\documentclass{elsart}
\usepackage{natbib}
\usepackage{amssymb}
\usepackage{graphicx,amsmath}
\usepackage{epstopdf}
\usepackage{bm}
\usepackage{color}

\usepackage{float}
\usepackage{threeparttable}

\newcommand{\bra}[1]{\langle #1|}
\newcommand{\ket}[1]{|#1\rangle}

\let\OLDthebibliography\thebibliography
\renewcommand\thebibliography[1]{
  \OLDthebibliography{#1}
  \setlength{\parskip}{0pt}
  \setlength{\itemsep}{0.0pt}
}
\journal{Advances in Atomic, Molecular and Optical Physics}
\begin{document}

\begin{frontmatter}

\title{Energy levels of light atoms in strong magnetic fields}

\author[1]{Anand Thirumalai}
\author[2]{Jeremy~S. Heyl}
\address[1]{School of Earth and Space Exploration, Arizona State University, Tempe, Arizona, USA 85287}
\address[2]{Department of Physics and Astronomy, University of British Columbia, Vancouver, BC, Canada V6T1Z1}

\date{\today}

\begin{abstract}
In this review article we provide an overview of the field of atomic structure of light atoms in strong magnetic fields. There is a very rich history of this field which dates back to the very birth of quantum mechanics. At various points in the past significant discoveries in science and technology have repeatedly served to rejuvenate interest in atomic structure in strong fields, broadly speaking, resulting in three eras in the development of this field; the historical, the classical and the modern eras. The motivations for studying atomic structure have also changed significantly as time progressed. The review presents a chronological summary of the major advances that occurred during these eras and discusses new insights and impetus gained. The review is concluded with a description of the latest findings and the future prospects for one of the most remarkably cutting-edge fields of research in science today. 
\end{abstract}


\end{frontmatter}

\tableofcontents

\section{\label{sec:intro}Introduction}
The field of atomic structure in strong magnetic fields is a truly remarkable and unique field of research. It has a rich and diverse history that dates back to the very foundations of quantum mechanics, to the late $19^{\textrm{th}}$ and early $20^{\textrm{th}}$ centuries. Just as remarkable as the field's longevity, is its prolificacy; so much so, that even at the time of writing this review, numerous computational atomic structure articles have appeared in the literature, making this review immediately incomplete. It is a virtually impossible task to include all the work that has gone into the field of atomic structure making it the fertile research landscape that it is today. The problem of atoms in magnetic fields is as remarkable in its classical severity as it is in terms of the beauty of its nuances, encapsulating at once, an entire branch of physics that evolved over decades, in a handful of simple equations that can be found in nearly every textbook of quantum mechanics today. This article will focus on a review of the most important works in atomic structure computations of light atoms in strong magnetic fields. This narrows the perspective considerably, yet incorporates all the salient features of the state-of-the-art in this field of research, finding application in a broad spectrum of areas as diverse as astrophysics, to materials science and chemical engineering, to atomic and molecular optics, to even pharmaceutical and biochemical sciences. 

The field of atoms in strong and intense magnetic fields ($B>10^9$~G\footnote{In terms of SI units $10^4$~G = 1T.}) is primarily a computational domain, since experiments are not possible in the current day. This is due to the fact that the strongest magnetic fields that can be sustained for any appreciable length of time in the laboratory are on the order of $10^5-10^6$~G in superconducting magnets, although recent strain experiments with graphene suggest that it is possible to create pseudo-magnetic fields of about $3 \times 10^6$~G. It is called a pseudo-magnetic field since the band structure of graphene is altered and partially flat bands can result at discrete energies, analogous to Landau levels \citep{Levy2010}, therefore the behavior of atoms is as though they are experiencing strong magnetic fields. However, in certain collider experiments actual transient magnetic fields in excess of $10^{18}$~G can be created for a fraction of a second \citep{Skokov2009}. However, these cannot be used for experiments aimed at determining atomic structure in strong magnetic fields. As a result, the only way of studying the structure of atoms in such magnetic fields is by means of theory and computation and the utilization of observations of perhaps the most wondrous astrophysical laboratories: neutron stars and magnetized white dwarfs that can routinely sustain the strongest magnetic fields present in the observable universe. We hope that this review of the work in this fundamental field of research will convey to the reader, a sense of the remarkable achievements made in this field and the directions in which developments are progressing today. 

\section{\label{sec:historical}Historical background}

Broadly speaking, the development of the field of atomic structure per se, can be characterized by three eras. The first \emph{historical era}, is characterized by perhaps the most momentous discoveries in quantum mechanics, which nearly every text in quantum mechanics contains. The story of atomic structure started during this era in 
$1927$, when one year after obtaining his doctorate, Douglas Rayner Hartree developed the self-consistent field method for atomic structure calculations \citep{Hartree1928} utilizing Schr\"{o}dinger's wave mechanics formulation, enabling approximate determination of the energies and wave functions of atoms and ions. A year later in $1928$ John Clarke Slater \citep{Slater1928} and John Arthur Gaunt \citep{Gaunt1928} showed that it would be possible to cast Hartree's original intuitive picture better by setting up a many-electron wave function for the atom as a product of one-electron wave-functions for the various electrons. Soon thereafter in 1930, \citet{Fock1930} and \citet{Slater1930} independently showed that using the Rayleigh-Ritz variational approach to small perturbations of the electrons' wave functions and requiring that the atom's energy remain stationary, it is possible to essentially derive the Hartree[-Fock] equations. This cast the entire method into a more rigorous framework, while still respecting the antisymmetrization requirement on the electrons imposed by the Pauli exclusion principle. Thereafter Hartree \citep{Hartree1935} extended his treatment to include a simpler prescription of Fock's original equations and a more practical and computationally tractable form of the Hartree-Fock equations emerged. The modern form of the Hartree-Fock equations can be written as,
\begin{eqnarray}
h\left(r_{i}\right)\psi_{i}\left(r_{i}\right) + \sum_{j\neq i}\left[\bra{\psi_{j}(r_{j})}w(r_{i},r_{j})\ket{\psi_{j}(r_{j})}\psi_{i}(r_{i})\right. \nonumber\\
\left.
-\bra{\psi_{j}(r_{j})}w(r_{i},r_{j})\ket{\psi_{i}(r_{j})}\psi_{j}(r_{i})\right]
=E_{i}\psi_{i}(r_{i}),
\label{eq:HF}
\end{eqnarray}  
where $h_i$ is the single particle hamiltonian which contains the kinetic and nuclear potential terms. Magnetic fields appearing in $h_i$ would contain both the linear and quadratic Zeeman terms (i.e. $\propto B$ and $\propto B^2$, respectively). $w(r_i,r_j) \propto e^2/|\vec{r}_i-\vec{r}_j|$ is the Coulomb interaction between the electrons. The first part of the second term in Eq.~(\ref{eq:HF}) is called the ``direct" interaction while the second part is called the ``exchange" which arises due to electron-spin. This latter term vanishes if the spins of the two interacting electrons ($\psi_i$ and $\psi_j$) are anti-aligned. These terms collectively represent the average Coulomb repulsion between electrons. The Hartree-Fock equations represent a coupled eigenvalue problem with a non-homogeneous term; the exchange between electrons. This coupling makes the problem analytically intractable, and also computationally intensive as the number of electrons increases. If the ``exchange" term is excluded then one obtains the Hartree equations, or ``equations without exchange". These equations established the foundation for carrying out atomic structure computations needed for investigating atoms in strong magnetic fields. In the following short review of important developments, for the sake of brevity, several notable contributions will regrettably need to be either glossed over or left unmentioned, and the review shall be streamlined towards atoms in strong magnetic fields.

Parallel to these developments, the first comprehensive explanation of the Zeeman effect in atoms came in $1939$ with two landmark studies by \citet{Jenkins1939} and by \citet{Schiff1939}, who respectively published experimental and theoretical treatises explaining accurately the quadratic Zeeman effect. It was also during this time that the importance of configuration interaction was becoming apparent in atoms, particularly for larger atoms with greater number of electrons \cite[][]{Green1940, Green1941}. 
 
From the very early stages, even as Hartree was formulating the so-called Hartree-Fock equations, it was realized that the energies calculated by the self-consistent field method had an inherent error associated with them on the order of $1-2\%$. The origin of this inaccuracy was well understood. The method of the self-consisent field assumes that the electrons move independently of one another and therefore only interact through averaged potentials of the other electrons. However, even from a classical perspective, it would be natural for the electrons to experience Coulomb repulsion from one another and therefore, any given electron would be less likely to be found in the vicinity of any other electron. Therefore, the idea was to account for this ``correlation" of the motion of various electrons.  The original idea for accounting for this correlation came from the brilliant work of Egil Andersen Hylleraas as early as in $1928$ \cite{Hylleraas1928} . He employed not a single determinental wave function, but rather a linear combination of determinants comprised of single-particle wave functions, forming a complete basis set. 
\begin{equation}
\Psi=\sum_{k=1}^{\infty}A_N\left(\psi_{1}^k, \psi_{2}^k, ..., \psi_{N-1}^k,\psi_{N}^k\right),
\label{eq:determinants_hylleraas}
\end{equation}
where $k$ denotes a certain configuration of electrons in the atom, and $A_N$ is the anti-symmetrization operator. The summation extends, in principle, over an infinite number of such configurations, thereby forming a complete basis set. The overlap integrals between the different spin-orbitals then accounted for the interaction between different configurations. \citet{Hylleraas1929} also suggested  that correlation could be handled in a much more intuitive manner by setting up, for helium, the ground state wave function to be a function of three independent variables; $r_1$ and $r_2$ the distances of the two electrons from the nucleus respectively, and $r_{12}$, the separation between them, with the latter expressing the correlation between the electrons. An explicitly correlated wave function could then be written as,
\begin{equation}
\Psi=\sum c_{l,m,n} s^l t^m u^n \textrm{exp}(-\alpha s),
\label{eq:hylleraas_three_vars}
\end{equation}
where $\{l,m,n\}$ are a set of three (non-negative) integers, the coefficients $c_{l,m,n}$ are variational coefficients to be optimized alongside a constant $\alpha$. The $\{s,t,u\}$ coordinate system is given by $s=r_1 + r_2$, $t=r_1 - r_2$ and $u=r_{12}$. Methods based on the latter technique yielded much faster convergence and accuracy, particularly for helium. These ideas were used extensively in the $1940$'s through to the $1960$'s yielding atomic structure for a variety of atoms with ever increasing accuracy. However, these treatises still only dealt with zero-fields and in some cases, magnetic fields of strength that were low enough that the interaction of the electron with the field was a small perturbation to their motion as largely dictated by the nucleus of the atom. Study of atomic structure in strong magnetic fields started off a new branch of study unto itself, but this would not occur until the mid-$1950$'s.  

\section{\label{sec:classical}The lightest `light' atom - hydrogen}
In $1956$, Yafet, Keyes and Adams \citep{Yafet1956}  investigated for the very first time, the effect of a strong magnetic field on the ground state of the hydrogen atom. While their motivation was to observe the effect in the case of impurities in semiconductors of high dielectric constants, their seminal work would mark the beginning of an altogether new era in the field of atomic structure in strong magnetic fields: \emph{``the classical era"} with motivations largely governed by solid-state applications. Their theoretical investigation consisted of increasing the magnetic field gradually from strengths in the perturbative regime to the strong field regime. In the former, the magnetic field is a perturbation to the motion of the electrons in the central field of the nucleus, while in the latter (at the infinite field limit, or the Landau regime) the nucleus is the perturbation to the interaction of the electron with the magnetic field. In the large intermediate range of magnetic field strengths in between, the problem is much more complicated as there is no suitable basis for expanding the wave function of the atom. \citet{Yafet1956} considered the Hamiltonian of the hydrogen atom in a uniform magnetic field to be given by,
\begin{equation}
H = -\nabla^2 + \gamma L_z + \frac{\gamma^2}{4}(x^2+y^2) - \frac{2}{\{x^2+y^2+z^2\}^{1/2}},
\label{eq:yafet_hamiltonian}
\end{equation} 
where the magnetic field strength parameter $\gamma$ is given by 
\begin{equation}
\gamma = \frac{\hbar \omega }{ 2 \mathrm{Ry}},
\label{eq:gamma}
\end{equation} 
$\omega=eB/m_e$ is the cyclotron frequency and Ry is the Rydberg energy. The second and third terms are the linear and quadratic Zeeman terms respectively, while the last is the central field of the nucleus. Correspondingly they employed a trial wave function with appropriate symmetries for the ground state of the hydrogen atom given by,
\begin{equation}
\psi=\left(2^{3/2} a_{\bot}^2 a_{\parallel} \pi^{3/2}\right)^{1/2} \exp\left(-\frac{x^2+y^2}{4a_{\bot}^2}+\frac{z^2}{4a_{\parallel}^2}\right).
\label{eq:yafet_wave_function}
\end{equation}
\begin{figure}
\begin{center}
\includegraphics[width=\textwidth]{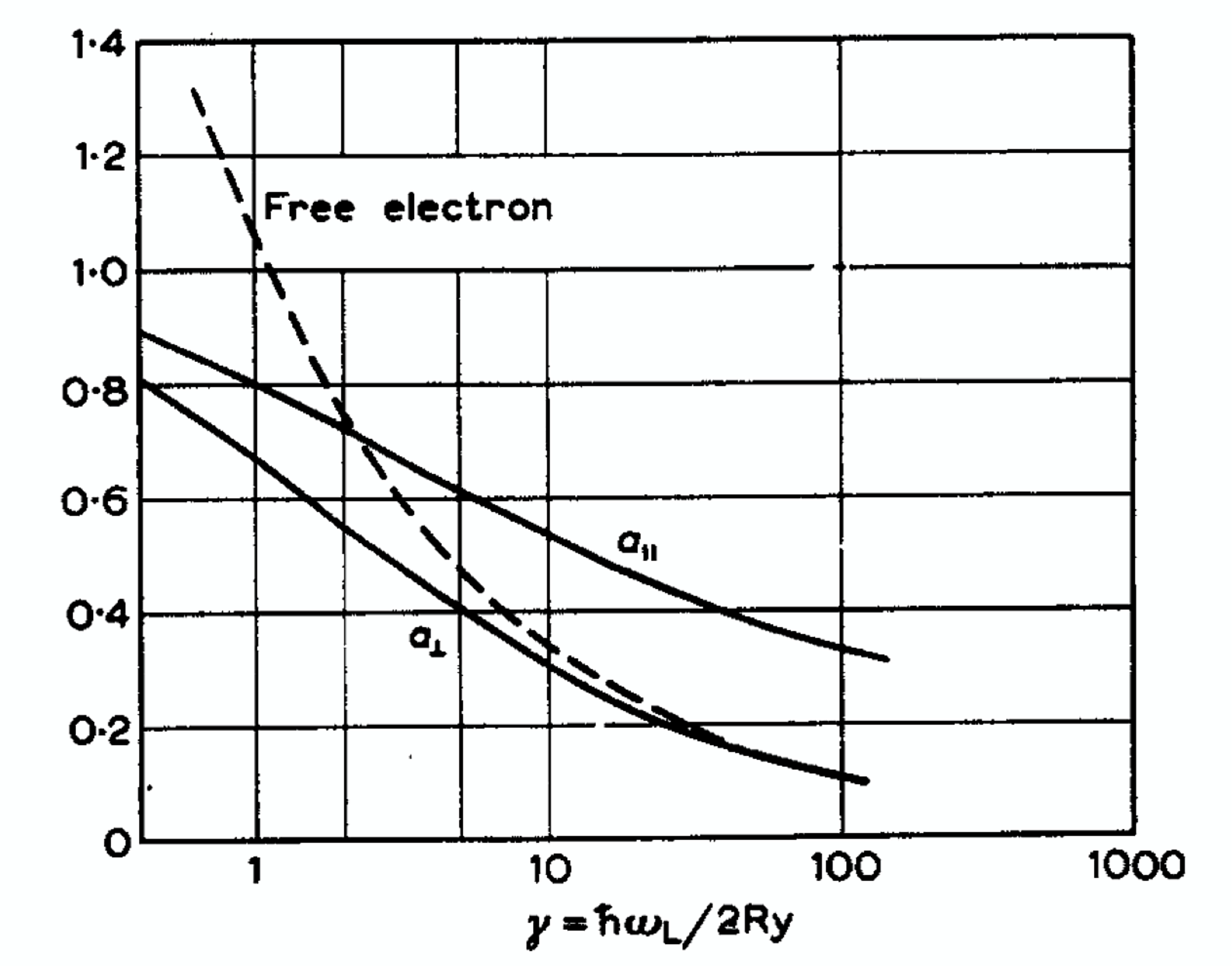}
\end{center}
\caption{Figure showing the shrinking dimensions of the hydrogen atom with increasing magnetic field strength. Notice that the atom shrinks in both directions parallel ($_\parallel$) and perpendicular ($_\bot$) to the magnetic field. Here $\omega_L$ is the cyclotron frequency. Figure reprinted with permission from~\citet[][]{Yafet1956} Copyright 1956 by Elsevier.}
\label{fig:yafet}
\end{figure}
The atom is placed in a uniform magnetic field pointing in the $z$-direction, with dissimilar dimensions of the atom in the parallel and perpendicular directions; $a_{\parallel}$ and $a_{\bot}$.
Using a variational approach and minimizing the ground state energy they obtained numerical solutions for the values of $a_\bot$ and $a_\parallel$ with varying magnetic field strength; see Figure~\ref{fig:yafet}. 

\begin{figure}
\begin{center}
\includegraphics[width=\textwidth]{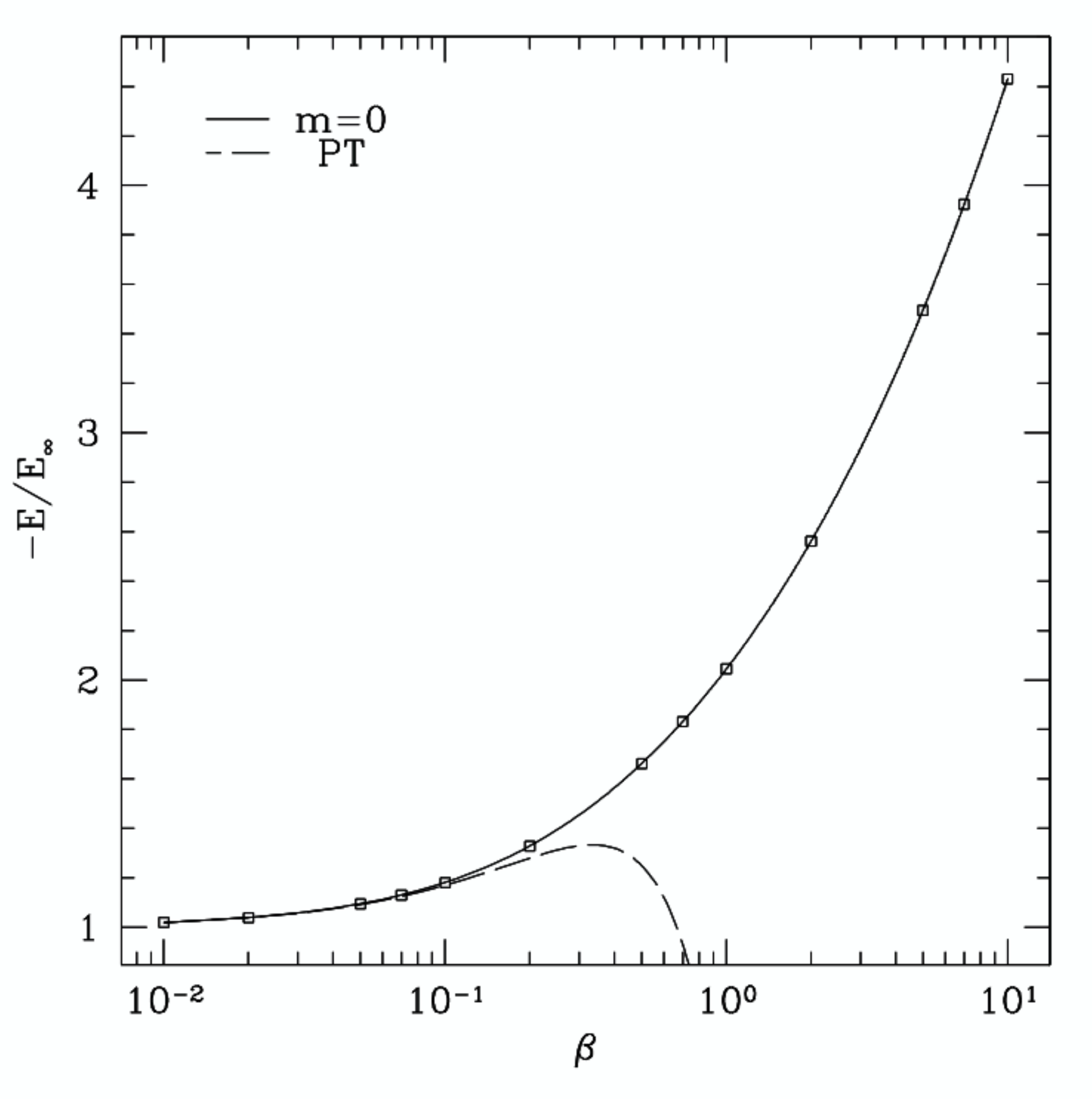}
\end{center}
\caption{Figure showing the breakdown of perturbation theory (PT dashed line) in the strong field regime,  reprinted with permission from~\citet{TH2009} Copyright 2009 by American Physical Society. Notice that the the binding energy $E$ of the ground state of the hydrogen atom (azimuthal quantum number $m=0$ corresponding to the field-free state $1S_0$) in strong magnetic fields increases approximately as the square of the logarithm of the field strength. $E_\infty$ is the Rydberg energy and the magnetic field strength parameter $\beta$ measures the magnetic field.}
\label{fig:PT_breakdown}
\end{figure}

They found that as the magnetic field increases, the hydrogen atom's ground state loses spherical symmetry, first becoming an egg-shaped ovoid in intermediate field strengths and later, cigar-shaped in the intense magnetic field regime. With increasing magnetic field strength they also found that the ground state of the atom became increasingly more bound, as the atom shrinks in all directions while simultaneously becoming elongated in shape along the magnetic axis. \citet{Yafet1956} were the first to consider a strong magnetic field in which perturbation theory breaks down; see figure~\ref{fig:PT_breakdown} which depicts the inadequacy of a perturbation theory calculation in the strong field regime, due to \citet{TH2009}. Also shown therein is the increasing binding energy $E$ of the ground state of the hydrogen atom (with azimuthal quantum number $m=0$) in the strong field regime as function of the strength parameter $\beta$ measuring the magnetic field $B$ in units of the reference magnetic field strength $B_0=2\alpha^2m_e^2c^2/(e\hbar) \approx 4.7 \times 10^{9}$~G with $\beta=\gamma/2$.

\begin{figure}
\begin{center}
\includegraphics[width=\textwidth]{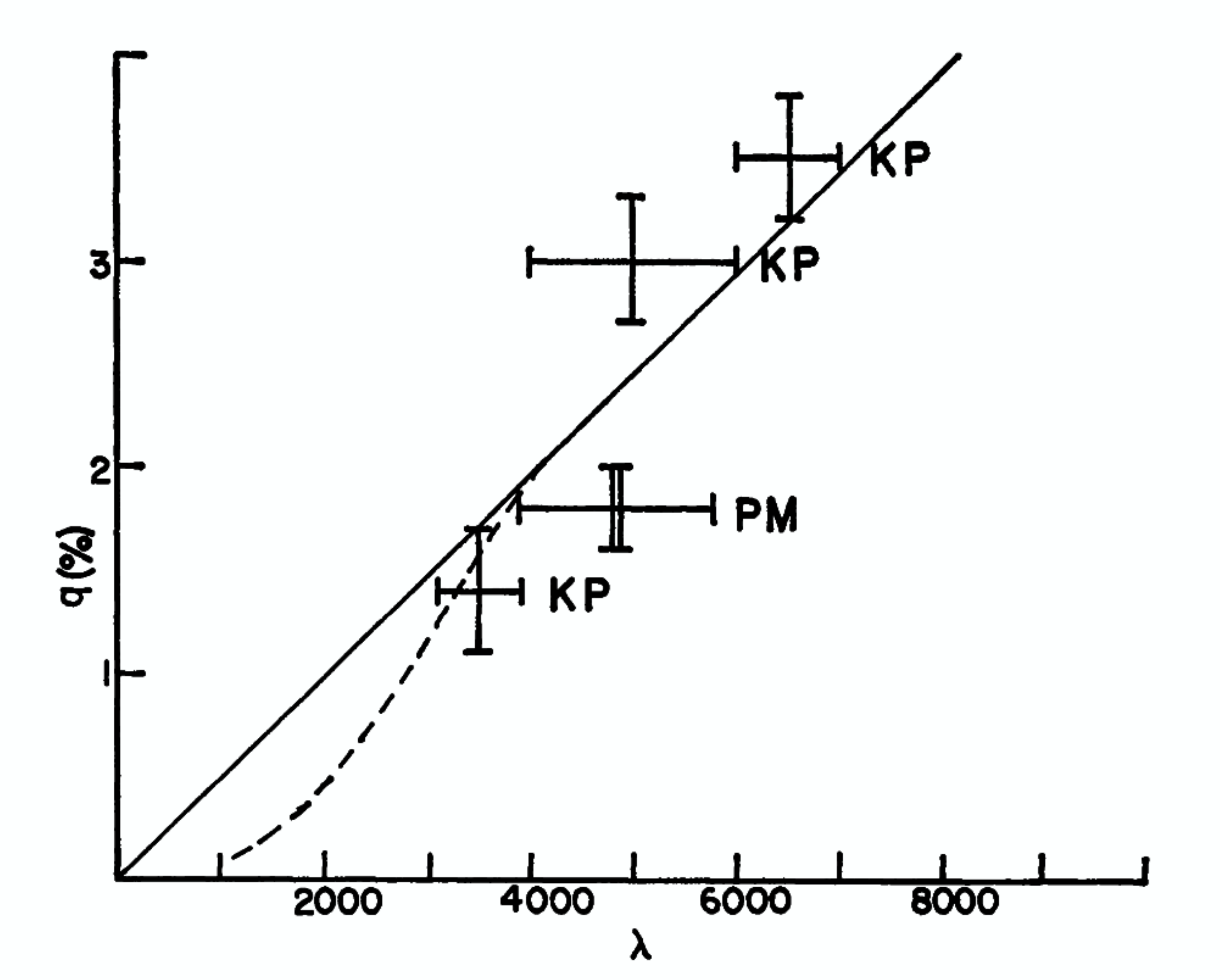}
\end{center}
\caption{Observations of \citet{Kemp1970} at Kitt Peak (KP) and Pine Mountain (PM)  observatories showing evidence for circular polarization from a magnetized white dwarf star. The quantity $q$ on the y-axis is the percentage of circular polarization.  The solid line shows grey body emission fit assuming a magnetic field of $1.2 \times 10^7$~G and the dashed line shows plasma effects. Figure from~\citet{Kemp1970}, Copyright 1970 AAS. Reproduced with permission.}
\label{fig:Kemp_mag_white_dwarf}
\end{figure}

From about $1950$ till about the end of the $1960$'s was a period of rapid growth in solid-state technologies. Advances in atomic structure theory were therefore leaning towards solid-state applications. It was only in $1961$ that \citet{Hasegawa1961} calculated the spectrum and oscillator strengths of the hydrogen atom in a uniform strong magnetic field and showed that in the limit of infinite field strengths, a simplified picture is obtained wherein the nucleus becomes the perturbation to the interaction of the electron with the field. This was the very first study to obtain the spectrum of hydrogen in strong magnetic fields. Subsequently, in the $1950$'s and $1960$'s there was a lot of interest in solid-state technologies and eventually this led directly to the development of density functional theory (DFT) in the mid-$1960$'s, by  \citet{Hohenberg1964} and \citet{Kohn1965}. Although including magnetic fields successfully in DFT was not achieved until $1987$ by \citet{Vignale1987}. This rapid growth in solid-state technologies was largely responsible for the increased sensitivity of astronomical polarimeters and as a result of such advances there came a momentous discovery that would rejuvenate interest in atomic structure in strong magnetic fields.  \citet{Kemp1970} observed strong circular polarization in the visible light from a ``peculiar" white dwarf. Until that time, it was theorized that white dwarfs may exhibit magnetism, but had not been observed. Their results were consistent with a magnetic field of about $10^7$~G in the white dwarf that they observed (see Fig.~\ref{fig:Kemp_mag_white_dwarf}).\\

Shortly thereafter, in a follow-up study \citet{Angel1971} observed similarly polarized light from a second white dwarf and within a decade it became well established that white dwarfs can harbor strong magnetic fields \citep[e.g.][]{Landstreet1975, Angel1978, Angel1981}. Even stronger magnetic fields were expected in the more exotic compact objects, neutron stars. However, discoveries of their magnetic fields had to wait until $1977-78$ when Tr\"{u}mper and co-workers \cite{Trumper1977, Trumper1978} discovered a strong line feature in the spectrum of the binary Her X-1, in which one of the stars is an accreting neutron star. They interpreted this as due to cyclotron emission and inferred a magnetic field of $4.6 \times 10^{12}$~G. This was the largest magnetic field observed in any star until that time. With the discovery of magnetized compact objects there occurred a shift in motivation for the study of atoms in strong magnetic fields, from solid-state physics to astrophysics, and the \emph{``modern era"} was ushered in.

As early as a year after the discovery of magnetized white dwarfs, motivated by astrophysical concerns, Riccardo~\citet{Barbieri1971} investigated the relativistic hydrogen atom in intense magnetic fields, characteristic of neutron stars, on the order of $10^{12} - 10^{13}$~G. By solving Dirac's equation he obtained an analytic expression for the ground state energy of the hydrogen atom in such field strengths. His work showed that the ground state binding energy increased with magnetic field, $B$ as,
\begin{equation}
E \sim \textrm{ln}(B/B_0)^2,
\label{eq:Barbieri}
\end{equation} 
with $B_0$ the reference magnetic field strength defined above.
\begin{figure}
\begin{center}
\includegraphics[width=\textwidth]{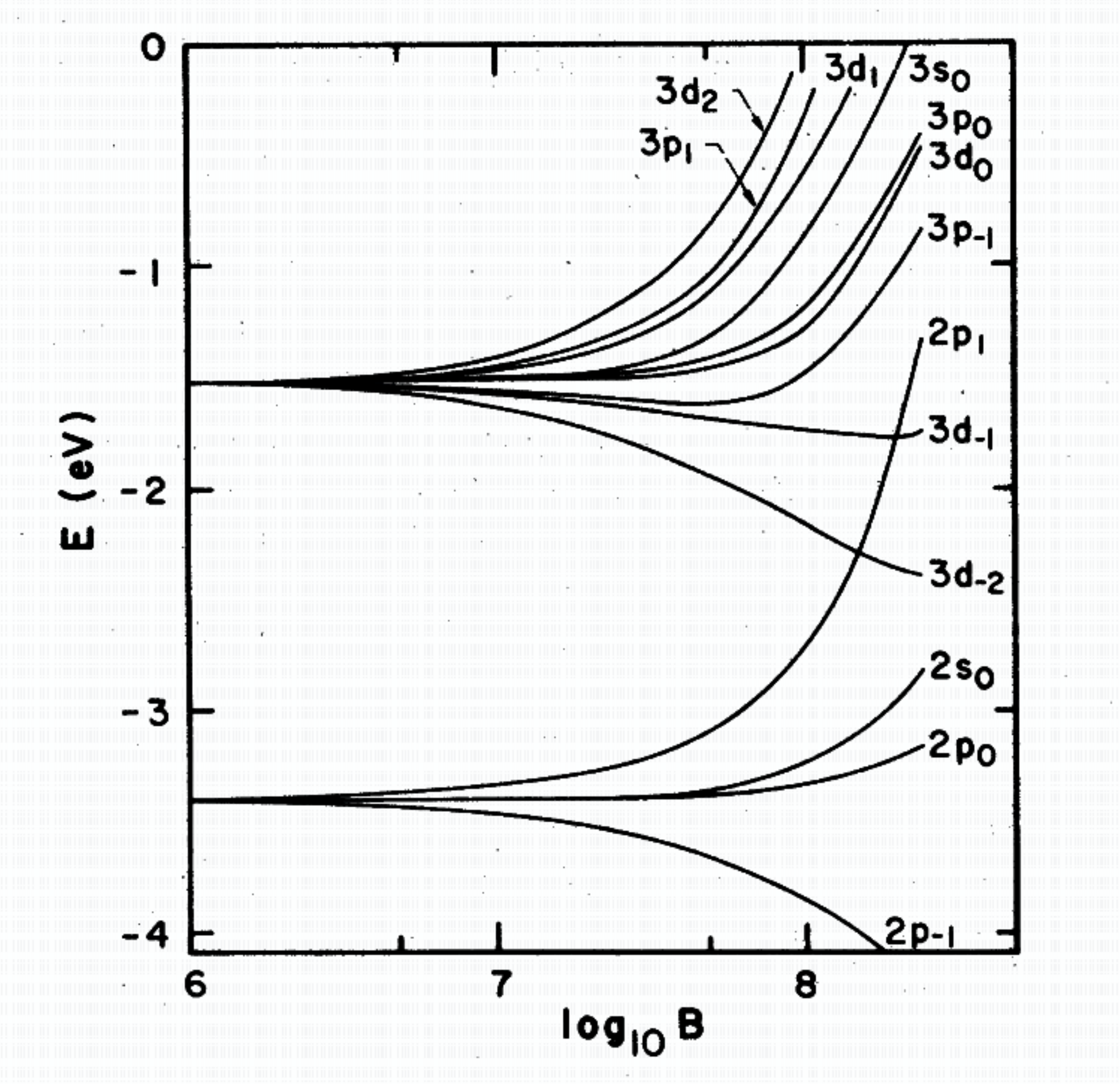}
\end{center}
\caption{Variation in the binding energies of the 13 lowest-lying states of hydrogen in strong magnetic fields, measured in G. Notice that the state with negative azimuthal quantum number, such as $2p_{-1}$, becomes more bound with increasing magnetic field strength, while other states with positive azimuthal quantum numbers such as $2p_1$ of the same triplet, show the opposite trend. Figure reprinted with permission from~\citet{Smith1972} Copyright 1972 by American Physical Society. }
\label{fig:Smith1972_fig2}
\end{figure}

Around the same time, in the early to mid-$1970$'s, Ed R.~Smith and co-workers \citep{Smith1972, Surmelian1974, Surmelian1976} determined the energy levels of about a dozen or so low-lying states of the hydrogen atom in strong magnetic fields, using a variational approach. They determined the behavior of the energy levels of 13 low-lying states of hydrogen with varying magnetic field strengths in the strong field regime (see Fig.~\ref{fig:Smith1972_fig2}).
They also determined bound-bound transition probabilities for the hydrogen atom \citep{Smith1973a, Smith1973b} to aid in atmosphere models of white dwarfs accounting for magnetic fields. Their efforts during this time represented the most comprehensive studies of the hydrogen atom in strong magnetic fields. There was also an effort by \citet{Hamada1973} to obtain estimates of binding energies for excited states of hydrogen using perturbation theory, but these were only applicable to about $2 \times 10^7$~G.  

\begin{figure}
\begin{center}
\includegraphics[width=12 cm]{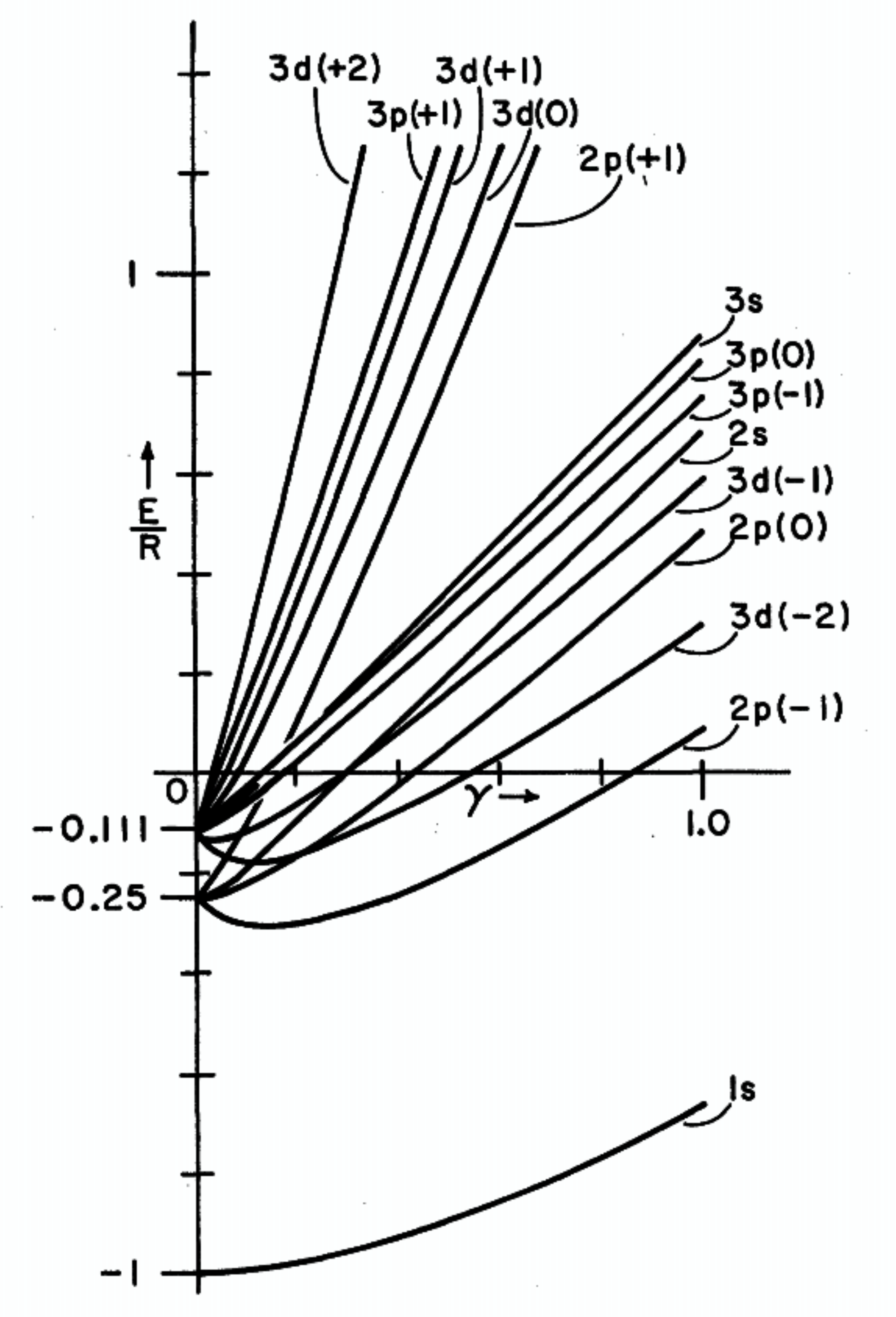}
\end{center}
\caption{Variation in the binding energies of the 14 lowest-lying states of hydrogen in strong magnetic fields in units of Rydberg energies.  Figure reprinted with permission from~\citet{Praddaude1972} Copyright 1972 by American Physical Society. }
\label{fig:praddaude_results}
\end{figure}

Parallel to these advancements, this period also saw some of the first fully numerical treatments of atoms in strong magnetic fields.  \citet{CK1972}  solved the problem of the hydrogen atom in the intense field regime using different approaches, including solving the one-dimensional Schr\"odinger equation numerically. The crux of their treatment was to utilize the adiabatic approximation, wherein the wave function of the electron separates into a product of two functions, one that is a function of $z$ alone, while the second which is a function of the remaining two orthogonal directions, as shown below,
\begin{equation}
\Psi = \sum_\alpha c_{\alpha} f_{\alpha}(z)\Phi_{\alpha}(x_1,x_2),
\label{eq:adiabatic}
\end{equation}
where $x_1$ and $x_2$ are the remaining two orthogonal directions, which could be $\{x,y\}$ in cartesian or $\{\rho,\phi\}$ in cylindrical coordinates. $\alpha$ is a set of quantum numbers and $c_{\alpha}$ a set of coefficients. In the adiabatic approximation the motion of the electron along the $z$-direction is not affected by the magnetic field. The orthogonal part of the wave function ($\Phi(x_1,x_2)$) can then be expanded using a set of Laguerre polynomials. Using such a wave function in the Schr\"odingier equation for the hydrogen atom in an intense magnetic field, they obtained a differential equation for solving for the unknown part of the wave function along the $z$-direction as,
\begin{equation}
\left[
\frac{\hbar^2}{2m} \nabla^2 + V_{ns}(z)
\right] f(z) = E~f(z).
\label{eq:canuto_kelly_diff_eq}
\end{equation}
The effective potential $V_{ns}$ is given by,
\begin{eqnarray}
V_{ns}(z) = \frac{4 \hbar c}{e^2} \sqrt{\frac{e \hbar B}{m_e^2 c^3}} \sum_{p=0}^n \sum_{q=0}^s \frac{(-)^{p+q}}{4^{p+q}} 
\begin{pmatrix}
n\\
n-p
\end{pmatrix}
\begin{pmatrix}
s\\
s-q
\end{pmatrix}
\frac{1}{p!q!} \times \nonumber\\
\frac{d^{2(p+q)}}{d\lambda^{2(p+q)}} \left[ e^{\lambda^2} \textrm{erfc}(\lambda)\right],
\label{eq:effective_pot}
\end{eqnarray}
where, $\textrm{erfc}(\lambda)=\int_{\lambda}^{\infty}e^{-x^2}dx$, is the complementary error function. Using this effective potential they solved the Schr\"odingier equation and obtained the binding energies of the ground and first few excited states. Elsewhere, H.C. ~\citet{Praddaude1972}, in the same year, established a new basis for expanding the wave function of hydrogen-like atoms in strong magnetic fields. He established a set of four quantum numbers (K,C,M,N) for describing the wave functions, similar to the canonical $n,l,m$ quantum numbers. He showed that this new basis given in Eq.~(\ref{eq:praddaude_basis}) which employed \emph{generalized} Laguerre Polynomials, reduced the Schr\"odinger equation to a set of algebraic equations which could be solved in an economical manner with relative ease, yielding binding energies for the 14 most low-lying states of hydrogen in strong magnetic fields. The wave function for the bound states defined as
\begin{equation}
\Psi = (2\pi)^{-1/2} \xi(\rho,z) \textrm{exp}(\textrm{i} M \phi),
\label{eq:praddaude_basis}
\end{equation}
can be expressed using generalized Laguerre polynomials as,
\begin{equation}
\xi(\rho,z) = z^C \rho^{|M|} e^{-|\gamma|\rho^2} e^{-2|\epsilon|^{1/2}r} \sum_{m=0}^N L_m^{|M|} (2 |\gamma| \rho^2) \times \sum_{k=0}^{\infty} \sum_{n=0}^{k} A_{mkn}\rho^{2n}L_{k-n}^{(\alpha)(4|\epsilon|^{1/2}r)},
\label{eq:praddaude_basis2}
\end{equation}
where,
\begin{equation*}
\epsilon=E/\mathcal{R} - \gamma M - |\gamma| (|M| + 2N + 1), 0 > \epsilon=-|\epsilon|,
\end{equation*}
\begin{equation*}
\alpha = 2(C + |M| +2N ) + 1,
\end{equation*}
\begin{equation*}
r=(\rho^2+z^2)^{1/2},
\end{equation*}
\begin{equation*}
C=0,1, ~~M=0,\pm1,\pm2, ... , ~~N=0,1,2, ... 
\end{equation*}
where $ \mathcal{R} = Z e^4 m_e/(32\pi^2\kappa^2 \hbar^2)$ is the effective Rydberg in a solid with dielectric constant $\kappa$. Figure~\ref{fig:praddaude_results} shows the variation in the binding energies of a few low-lying states of hydrogen as a function of the magnetic field, as obtained by Praddaude using this specialized basis. \\

Through the mid-$1970$'s there was a considerable amount of work in determining with ever increasing accuracy the energy levels of hydrogen in strong and intense magnetic fields and Roy Garstang's excellent review of ``atoms in high magnetic fields", published in $1977$ \citep{Garstang1977}, represents a summation of all the work done up to that point, motivating further research in high magnetic field atomic structure from a spectroscopic standpoint. 

\begin{figure}
\begin{center}
\includegraphics[width=\textwidth]{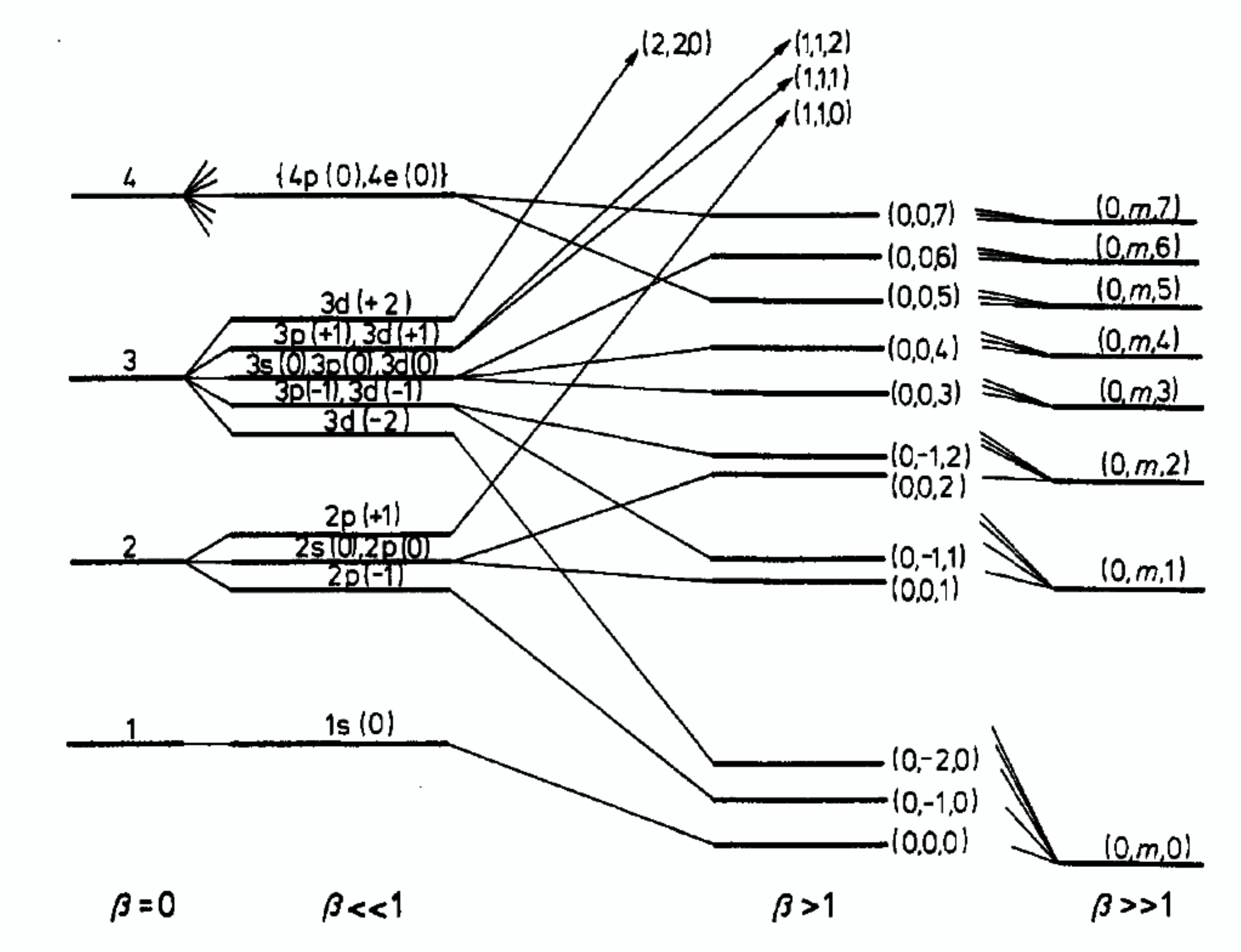}
\end{center}
\caption{The correspondence diagram between field-free and strong-field eigenstates. The quantum numbers $(n,m,k)$ given in parentheses count the nodes in the orthogonal directions $\rho, \phi$ and $z$, respectively, with $n=0$ giving the ground Landau level. Figure from~\citet{SV1978} Copyright 1978 IOP Publishing. Reproduced with permission. All rights reserved. }
\label{fig:correspondence}
\end{figure}

A year later, \citet{SV1978} approached the problem numerically from a different angle. They began at the infinite field limit with an expansion of the wave function using Landau orbitals, and as they then approached the finite field case by reducing the magnetic field strength, the Coulomb coupling became more appreciable and they obtained a set of coupled differential equations for solving for the unknown part of the wave function along the magnetic axis. They expanded the wave function in the adiabatic approximation as, 
\begin{equation}
\Psi=\psi(z) \left(\frac{eB}{2 \pi \hbar}\right)^{1/2} \exp(\textrm{i} m \phi) \exp(-\zeta/2)\zeta^{|m|/2} P_{nm}(\zeta),
\end{equation}
where $\psi(z)$ is the unknown part of the wave function along the magnetic axis, $\zeta=\rho^2 e B / 2\hbar$, and the orthogonal part of the wave function consisting of Landau orbitals with the polynomials $P_{nm}$ being closely related to the associated Laguerre polynomials according to,
\begin{equation}
P_{nm}(\zeta) = \frac{1}{(n!s!)^{1/2}}\sum_{k=0}^{\textrm{min}(n,s)} \frac{(-1)^k}{k!}
\begin{pmatrix}
n\\
n-k
\end{pmatrix}
\begin{pmatrix}
s\\
s-k
\end{pmatrix}
\zeta^{\textrm{min}(n,s)-k}~~~(s \equiv n-m).
\end{equation}
This was the first time the problem had been approached numerically from the infinite field limit and their study revealed some very important nuances. First, they found that there existed some altogether new correspondences between the field-free $(n,l,m)$ state and strong-field eigenstates $(n,m,k)$, correcting errors that other researchers had made up to then. The quantum numbers in the strong-field case count the nodes in the orthogonal directions, $\rho, \phi$ and $z$ respectively. Figure~\ref{fig:correspondence} shows the correspondence between the different states as a function of magnetic field strength. Second they found that not all eigenstates are bound states, even though they appear as such in the adiabatic approximation. They found that several of these metastable states would make a radiation-less transition to a free state.\\ 

 Although Simola and Virtamo's work produced the most accurate results up to that time, there was a limitation that it was not accurate for highly excited states in strong magnetic fields. This difficulty was overcome by Helmut~\citet{Friedrich1982} by solving for the spectrum of hydrogen by going beyond the adiabatic approximation using a non-orthogonal basis which separates the Landau orbitals into functions of the constituent variables using displaced gaussians. This ultimately produced a coupled eigenvalue problem in the form of an ordinary differential equation with coupling between different channels in the expansion. He solved this using a diagonalization method and found that his overall methodology made it possible to accurately determine the binding energies of highly excited states, which was not possible until then. By the late $1970$'s and early $1980$'s, efforts with the hydrogen atom were rapidly becoming computationally complex and there was a growing concern regarding reproducibility, given the fact the these computed wave functions were not easily available at the time. Additionally, not every researcher had at his disposal computing infrastructure that could handle the computational requirements imposed by such methods as those of Simola and Virtamo and Friedrich. Motivated by a very genuine concern to make these computations tractable using standard integration and diagonalization routines at the disposal of the average researcher, \citet{Baye1984} devised a simple variational basis  that was not only accurate but also easy to handle numerically speaking,
\begin{equation}
\psi^{m}_{\alpha_i \beta_j} = \rho^{|m|} \textrm{exp}(\alpha_i \rho^2) \textrm{exp}(\textrm{i} m \phi) \textrm{exp}(-\beta_j |z|),
\label{eq:baye_vincke}
\end{equation}
where the parameters $\alpha_i$ and $\beta_j$ could be optimized in a variational calculation yielding accurate results.

Elsewhere, during the two decades leading up to the 1980's, one of Hartree's students Charlotte Froese-Fischer, led the development of some of the first sophisticated multi-configuration Hartree-Fock atomic structure calculations of the time. These calculations were a significant milestone in atomic structure, as they were able to run on computing architecture prevalent at the time, using portable algorithms written in FORTRAN. Eventually, in 1977 she published a book,  \emph{The Hartree-Fock method for atoms: a numerical approach} \cite[][]{CFF1977}, which represented the state-of-the-art in atomic structure theory and computations. Her calculations had matured to the point that accurate structure of atoms from hydrogen to radon could be computed with effects such as electron correlation included along with relativistic and other accompanying corrections as well as electron screening for the larger atoms. 

\begin{figure}[H]
\begin{center}
\includegraphics[width=\textwidth]{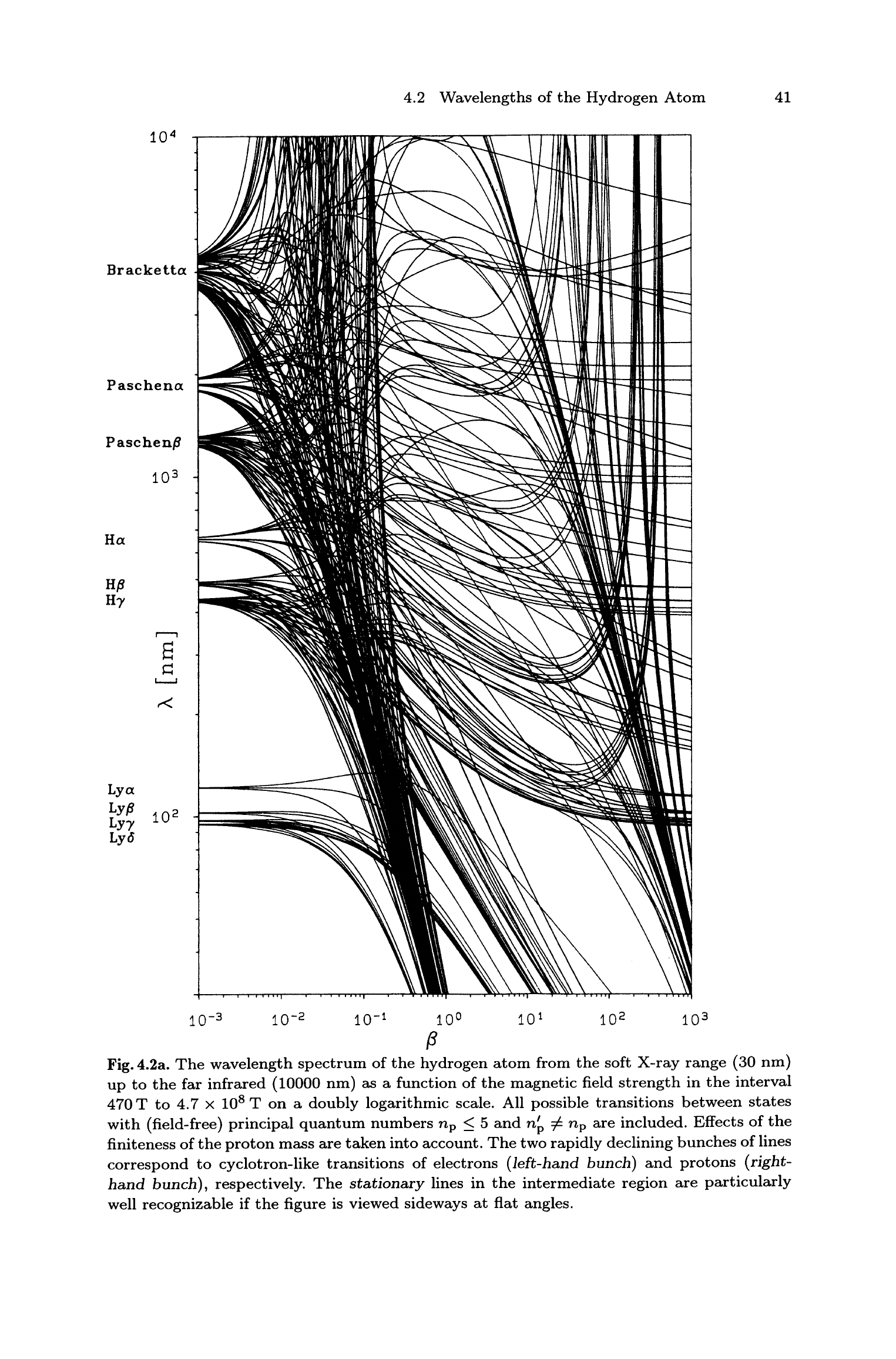}
\end{center}
\caption{The wavelength spectrum of the hydrogen atom in a range of magnetic field strengths. The emission wavelength, $\lambda$ in nm, vs the magnetic field strength parameter $\beta=\gamma/2$ on the lower scale. Figure from~\citet{Wunner1987} Copyright 1978 IOP Publishing. Reproduced with permission. All rights reserved.}
\label{fig:rosner_atoms}
\end{figure}

By $1982$, her code could be adapted for tackling the problem of atoms in strong magnetic fields and \citet{Wunner1982} utilized wave functions computed using her code for determining the energies and energy-weighted sum rules for electromagnetic dipole transitions in hydrogen-like atoms in arbitrary field strengths. In the same year, they were also able to utilize Froese-Fischer's code for determining the structure of helium, as well as later for positively charged ionic species with two electrons in a whole range of magnetic field strengths (see below). By the mid-$1980$'s the hitherto most comprehensive list of energies and transition wavelengths for hydrogen in strong and intense magnetic field strengths had emerged \citep{Rosner1984, Wunner1985}, which R\"osner and Wunner and co-workers utilized to analyze the spectrum of a magnetized white dwarf. Figure~\ref{fig:rosner_atoms} shows their beautiful results for the hydrogen atom showing how the different transition wavelengths change with varying magnetic field strength. This was a major milestone in atomic structure in strong magnetic fields and encapsulated about thirty years of cumulative work in the scientific community. Their efforts during the 1980's and early 1990's culminated in their book which represents, even today a standard reference for atomic structure in strong magnetic fields \citep{Ruder94}. 

\section{\label{sec:few}Light atoms: two and few-electron systems}

Parallel to the development of methods aimed at determining the structure of hydrogen in strong magnetic fields, there was also a considerable amount of effort dedicated towards helium. With regards to few-electron systems however, there is very little data available in the literature, even to this day. 

One of the very first studies to investigate the structure of light atoms in strong and intense magnetic fields was as early as $1970$, by Cohen, Lodenquai and Ruderman \cite[][]{Cohen1970}. Using a purely variational approach with a few variable parameters they were able to arrive at initial estimates for the ground state binding energies of a handful of atoms; hydrogen, helium, lithium, boron and neon. This was nearly a decade before the confirmation of strong magnetic fields being present in neutron stars. Around the same time, \citet{Surmelian1973} calculated the energy spectrum of neutral helium in strong and intense magnetic fields computing data for the ground and first 13 excited states as well as bound-bound transition probabilities in magnetic fields of $10^7-10^9$~G. Once again their approach was a purely variational one with the wave function comprised of spherical harmonics and a radial part, which consisted of a combination of power law and exponentials to be optimized. A contemporary PhD student of Surmelian at the time, R.O. Mueller, along with co-workers A.~R.~P.~Rau and Larry Spruch, carried out variational calculations \citep{Mueller1975} in the same vein, obtaining variational upper bounds for the energies of a few two-electron systems such as $\textrm{H}^-$, He, and $\textrm{Li}^+$. There was even an effort by Banerjee, Constantinescu and Reh\'ak \citep{Banerjee1974} to arrive at rudimentary estimates for the energy levels of atoms using a statistical approach; a Thomas-Fermi model for atoms in strong magnetic fields. There were also efforts by Glasser and Kaplan \citep{Glasser1975, GK1975} to determine the structure of condensed matter; \textbf{a chain of atoms in the crust} of a neutron star with a strong magnetic field. While in the atmospheres of these compact objects isolated atoms are energetically favored, such may not necessarily be the case in their highly magnetized crusts. \citet{Ruderman1971} found that in the case of a neutron star's crust, condensed matter likely takes the form of linear chains of atoms and molecules, with each chain surrounded by a sheath of electrons. \citet[][]{GK1975} were motivated by the need to include electron correlation into Ruderman's model. In this picture, an understanding of solitary atoms in strong magnetic fields therefore plays a central role for understanding the nature of condensed matter in the same. In the latter case, the electrons interact with not one nucleus but rather a chain of them. The other interactions in this case include the inter-electron interactions including exchange, as well as interactions between the different nuclei themselves. Thus, understanding of electron-electron and electron-nucleus interaction in the case of solitary atoms forms the basis for extending the treatment to the case of chains of atoms or nuclei. It is possible to treat the latter case in the Hartree-Fock approximation as well \citep[e.g.][see below]{Neuhauser1987}. In their early work however, \citet{Glasser1975} and \citet{GK1975} studied the nature of inter-electron interactions in such condensed matter in strong magnetic fields using a purely variational approach, and found that inter-electron repulsion leads to the formation of anisotropic crystalline structure. This results partially because the motion of the electrons is not constrained in the direction parallel to the magnetic field, but is severely constrained in the transverse direction \citep[e.g.][]{Neuhauser1987}.

\begin{figure}[h]
\begin{center}
\includegraphics[width=\textwidth]{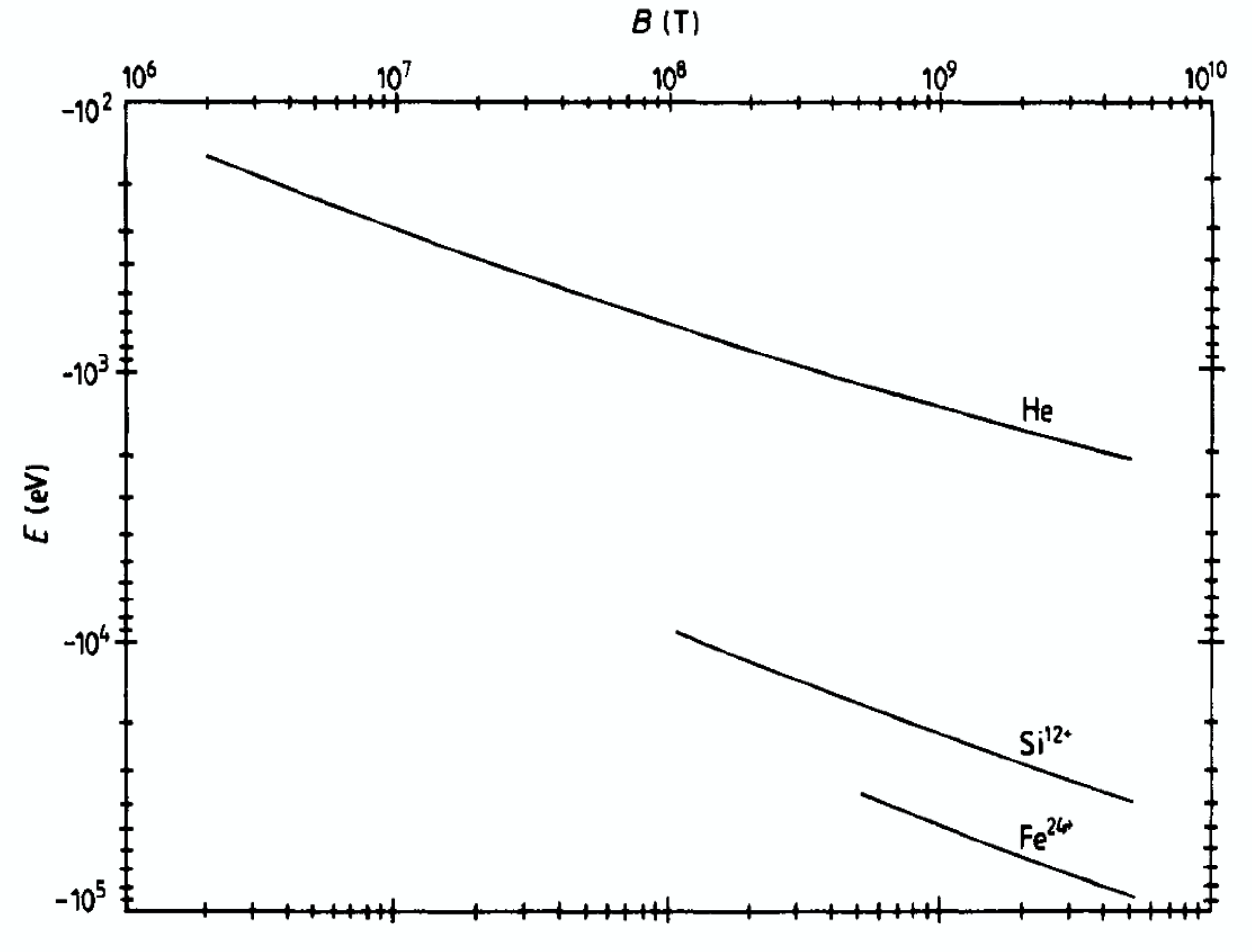}
\end{center}
\caption{The ground state energies of He and the He-like ions, Fe$^{24+}$ and Si$^{12+}$ as a function of magnetic field strength. Figure reprinted with permission from~\citet{Proschel1982} Copyright 1982 by the IOP Publishing. Reproduced with permission. All rights reserved.}
\label{fig:proschel}
\end{figure}

The advent of portable numerical routines alongside growth in computing infrastructure during the late $1970$'s and early $1980$'s provided further impetus for numerical efforts at determining the structure of atoms in strong magnetic fields. Pr\"oschel and co-workers \citep{Proschel1982} utilized the by then robust Hartree-Fock computer codes of Charlotte Froese-Fischer, with heavy modifications, to determine the energy levels of low-lying states of helium atoms in strong magnetic fields, in the adiabatic approximation. Their computation was based upon expanding the wave function of helium using Landau orbitals in the $\rho$ and $\phi$ directions in cylindrical coordinates and then solving for the unknown part of the wave function along the $z$ direction. They were able to provide binding energies of several low-lying states of helium and this study represented one of the first fully numerical Hartree-Fock computation of atoms in strong magnetic fields. They were also able to provide ground state energies of He-like ionized systems, up to nuclear charge $Z=26$, in magnetic fields relevant for neutron stars, see Fig.~\ref{fig:proschel}.
 
\begin{figure}[H]
\begin{center}
\includegraphics[width=\textwidth]{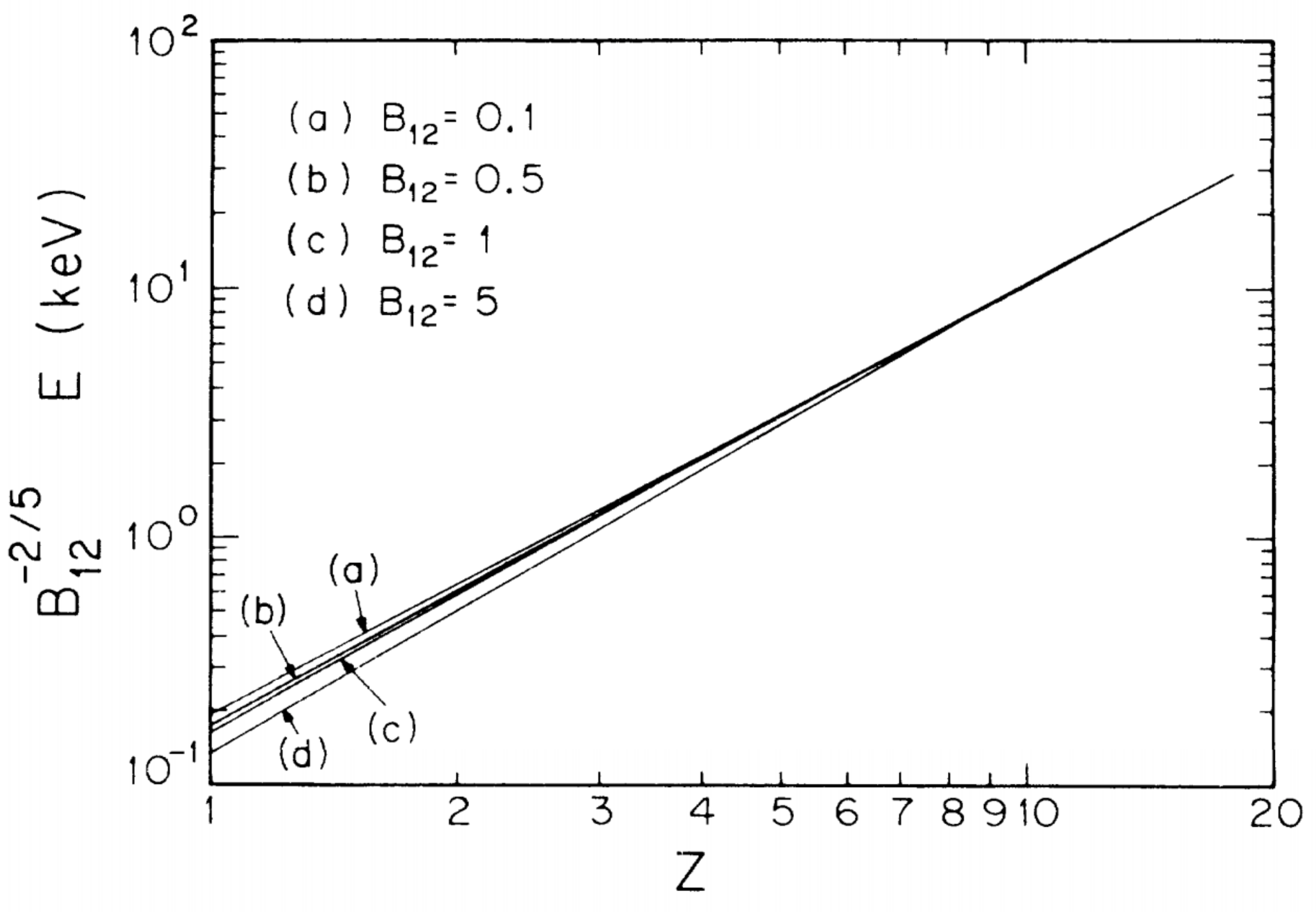}
\end{center}
\caption{The ground state energies of atoms up to $Z=18$ as a function of magnetic field strength {with $B_{12}=B/10^{12}$~G. } 
 Figure reprinted with permission from~\citet{Neuhauser1987} Copyright 1987 by the American Physical Society.}
\label{fig:neuhauser}
\end{figure}

The first study however, to investigate condensed matter heavier than helium in intense magnetic fields with the correct representation of exchange between electrons was by Neuhauser, Langanke and Koonin \citep{Neuhauser1986, Neuhauser1987}. They considered a chain of nuclei with equal spacing with the $Z$ electrons per unit cell being confined to Landau orbitals by the magnetic field, with motion along the chain governed by electrostatic interactions with and between other nuclei and electrons. Their Hartree-Fock calculation revealed that for atoms with $Z>2$, isolated atoms are energetically favored over molecular chains on the surface of neutron stars, in contrast to earlier calculations. They were also able to calculate the ground state binding energies of atoms up to $Z=18$ and derived an empirical scaling relationship for the binding energies as,
\begin{equation}
E \sim -158~\textrm{eV}~\times~Z^{9/5}\left (\frac{B}{10^{12}~\mathrm{G}}\right )^{2/5},
\end{equation}
estimated from their results for isolated atoms (see Figure~\ref{fig:neuhauser}).

Prior to the exact treatment that allowed magnetic fields to be accounted for successfully in DFT due to \citet{Vignale1987}, \citet{Jones1985, Jones1986} as well as \citet{Kossl1988} calculated the ground state binding energies of atoms, molecular chains and solids in lattice form, on the surface of neutron stars with intense magnetic fields. They however had to work within the limitations of DFT at the time, namely that exchange and correlation was only approximately accounted for with errors therein. In addition, most of these computations were still only restricted to the ground state configurations.

\begin{figure}[H]
\begin{center}
\includegraphics[width=\textwidth]{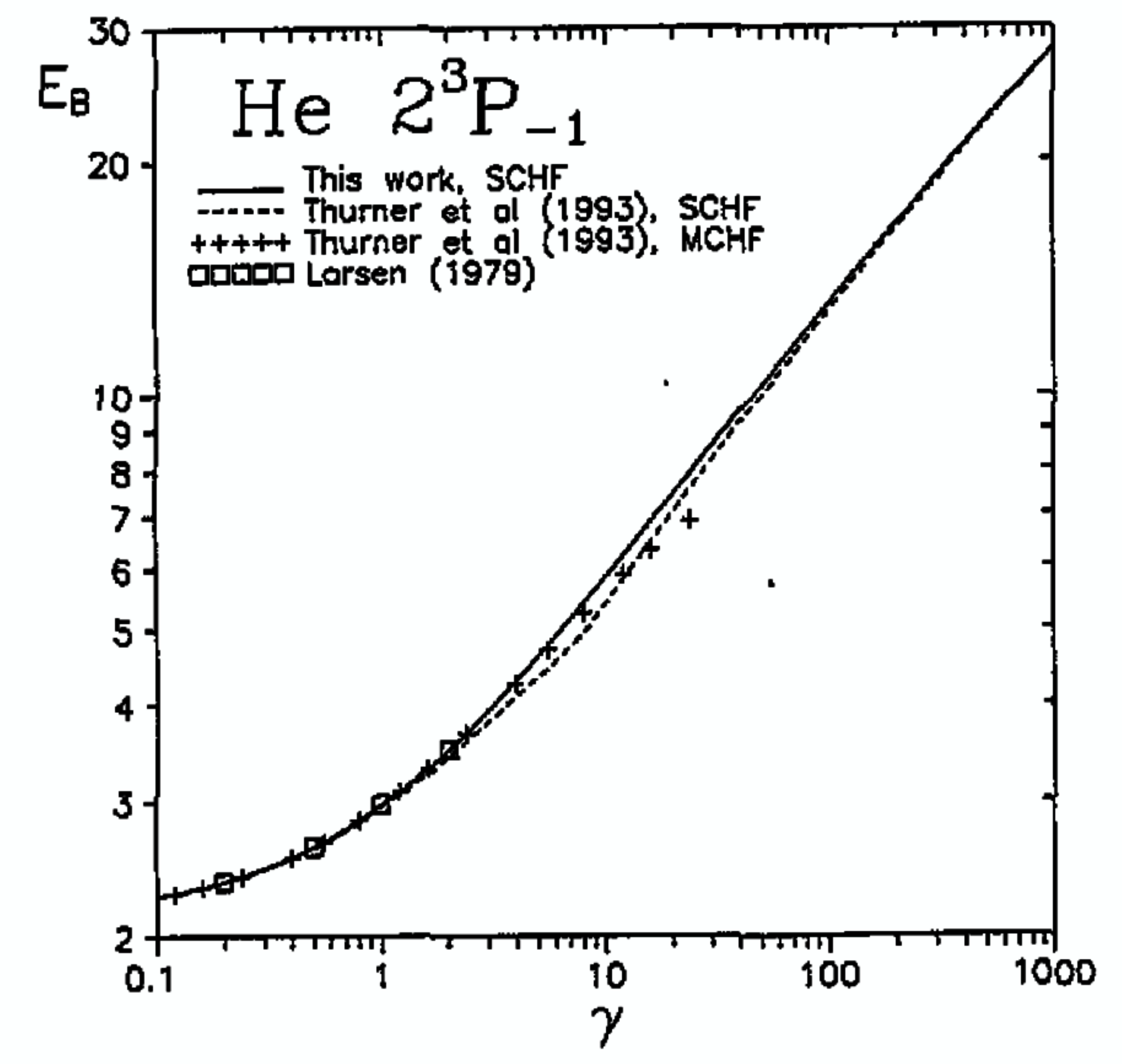}
\end{center}
\caption{The binding energy ($E_B$ [in atomic units]) of the ground state of helium as a function of magnetic field strength parameter ($\gamma$). The two-dimensional calculation was more accurate than the previous one-dimensional counterparts by \citep{Thurner1993,Larsen1979}. Figure from~\citet{Ivanov1994}. Copyright 1994 IOP Publishing. Reproduced with permission. All rights reserved. }
\label{fig:ivanov}
\end{figure}

By the mid-1990's, spectra of magnetized white dwarfs were commonplace. It was also now possible due to fast computer architectures to carry out Hartree-Fock and DFT computations with more ease than ever before and a great wealth of data began to emerge. By this time, the binding energies of the majority of the low-lying states of helium, as well as oscillator strengths were known reasonably accurately, in strong and intense magnetic fields. Progress therefore occurred essentially in two simultaneous directions. First, computations began to emerge for the hydrogen molecule accounting for electron correlation using a multi-configuration approach using the self-consistent Hartree-Fock technique, albeit in one-dimensional form \citep{Miller1991, Lai1996} and second, the problem of atoms in strong fields was cast into a two-dimensional form by \citet{Ivanov1988, Ivanov1994}. Ivanov's works were the first studies to approach the problem as a two-dimensional one. An atom in a magnetic field only has one predominant symmetry, namely azimuthal symmetry, if the magnetic field is aligned along the $z-$direction. Utilizing this natural symmetry, the problem can be expressed in three-dimensional form in cylindrical coordinates as,
\begin{equation}
\Psi=\psi(\rho,z)e^{-\textrm{i}m\phi}.
\end{equation}
The key advantage was that the wave function was not restricted to the adiabatic approximation. After integration in the $\phi-$direction the resulting Hartree-Fock equations then take on a coupled partial differential form in two dimensions,
\begin{eqnarray}
H_i \psi_i(\rho,z) + \left[\sum_{j \neq i}J_j(\rho,z)\right]\psi_i(\rho,z) - \left[\sum_{j \neq i}K_j(\rho,z)\right]\psi_i(\rho,z) = \nonumber\\
 \epsilon_i \psi_i(\rho,z),
\end{eqnarray}
where $J_j$ and $K_j$ are the direct and exchange kernels determined using estimates of the wave functions from the previous iteration. The single particle Hamiltonian is given by,
\begin{equation}
H_i=-\frac{1}{2}\left(\frac{\partial^2}{\partial \rho^2} + \frac{1}{\rho}\frac{\partial}{\partial \rho} + \frac{\partial^2}{\partial z^2} - \frac{m_i^2}{\rho^2} \right) + \left(s_{z,i} + \frac{m_i}{2}\right)\gamma + \frac{\gamma^2}{8}\rho^2 - \frac{Z}{\sqrt{\rho^2+z^2}},
\end{equation}
where $\gamma$ is the magnetic field strength parameter defined in Eq.~\eqref{eq:gamma}. Ivanov determined the binding energies of the first few low-lying states of hydrogen \citep{Ivanov1988} and helium \citep{Ivanov1994} using this approach. His investigation showed that this prescription resulted in binding energies that were more accurate than those obtained by earlier investigations (see Figure~\ref{fig:ivanov}). However the problem now became computationally far more intensive than its one-dimensional counterpart.

\begin{figure}[H]
\begin{center}
\includegraphics[width=0.7\textwidth]{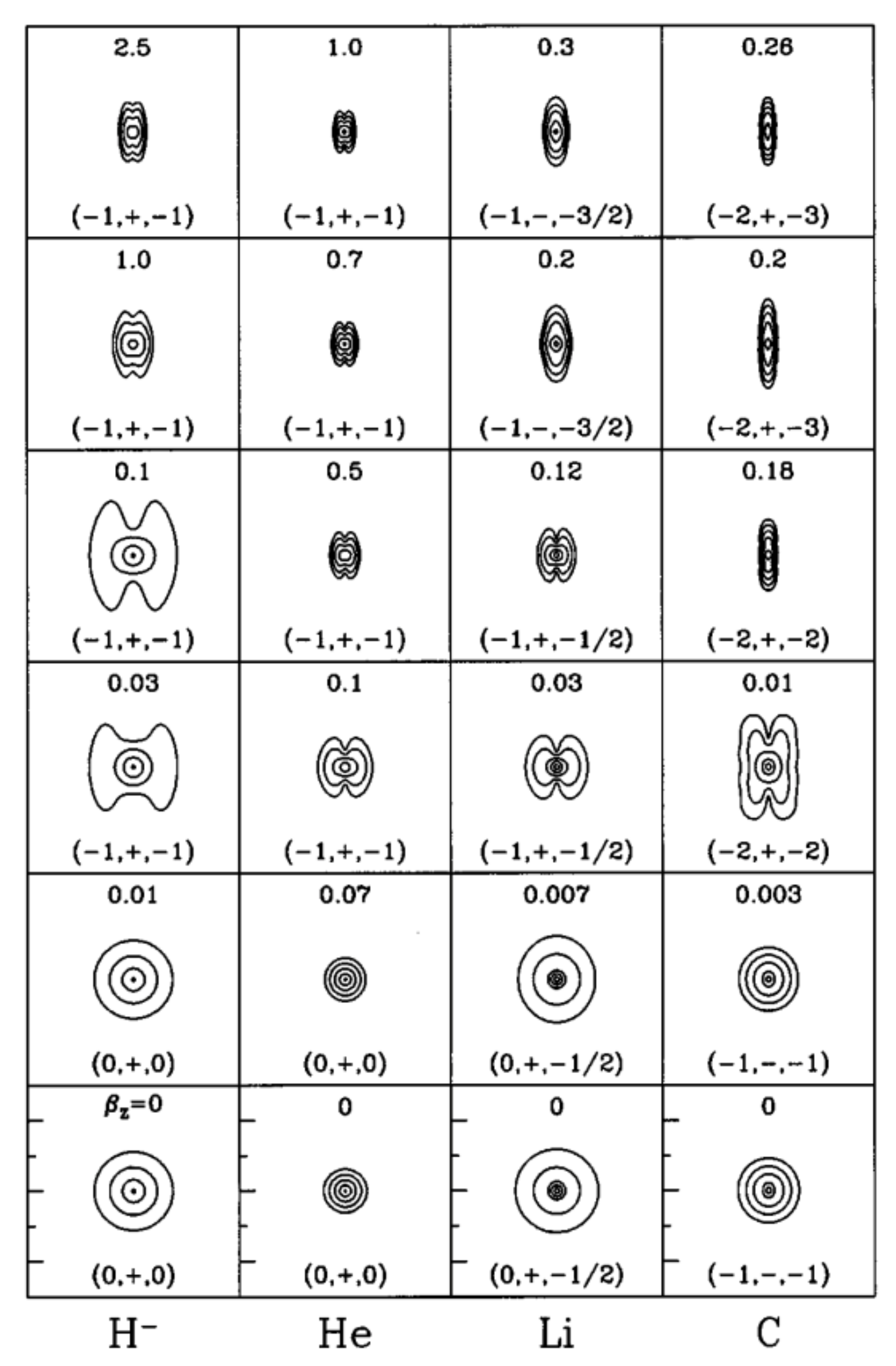}
\end{center}
\caption{Electron densities of some low-lying states of H$^-$, He, C and Li. The quantum numbers are $(M, \Pi_z, S_z)$; the total azimuthal quantum number, parity and $z$-component of spin. Notice the breakdown of spherical symmetry with increasing magnetic field strength. Here $\beta_Z=\gamma/2Z^2$ is the magnetic field strength parameter. Figure reprinted with permission from~\citet{Jones1996}. Copyright 1996 by the  American Physical Society.}
\label{fig:jones}
\end{figure}

A separate direction was taken by Jones et al \citep{Jones1996, Jones1997, Jones1999} in the late $1990$'s. They utilized different quantum Monte Carlo
methods (QMC) including ``released-phase" QMC which also allowed them to extend the correlation function method to complex Hamiltonians and wave functions, enabling estimation of excited state energies. The crux of the idea behind quantum Monte Carlo techniques is to utilize a random walk to sample a multi-dimensional space in which integrals are computed. These integrals are typically expectation values of different observables, say the system's energy or particle momentum for example. Such integrals become rapidly intractable to solve using regular quadratures with growing number of particles, which is the point where Monte Carlo methods for evaluating multi-dimensional integrals become useful. However to do so, a sufficiently good starting guess for the unknown many-body wave function is required. Using such a method Jones and co-workers were able to arrive at very accurate estimates for the binding energies of several low-lying states of helium \citep{Jones1997, Jones1999} as well as other low-$Z$ atoms such as Li and C \citep{Jones1996}. Their results for two- and few-electron systems are shown in Figure~\ref{fig:jones}, which shows how atoms undergo breakdown of spherical symmetry with increasing magnetic field strength. 
However, despite the underlying simplicity of the technique, the approach still required a significant computational overhead, particularly for greater number of electrons. 

While the majority of the studies up to this point concerned themselves with strong and intense magnetic fields up to about $10^{12}$~G, very few studies had investigated the regime between about $10^{12} - 10^{15}$~G, which can be found in highly magnetized neutron stars $-$ magnetars. One of the authors (JSH) of the current article, in $1998$ investigated the problem of the hydrogen atom and molecule as well as the helium atom in intense magnetic fields upwards of $10^{11}$~G \cite[][]{HH1998}. They employed the adiabatic approximation in which they expressed the wave function as,
\begin{equation}
\psi_{0m\nu}=R_{0m}(\rho,\phi)Z_{m\nu}(z)\chi(\sigma),
\end{equation}
where
\begin{equation}
R_{0m}(\rho,\phi)=\frac{1}{\sqrt{ 2^{|m|+1} \pi |m|! } a_H^{|m| + 1}}\rho^{|m|}\exp\left (-\frac{\rho^2}{4a_H^2} \right)e^{im\phi},
\end{equation}
with $a_H=\sqrt{\hbar c / (e B)}$ and $\chi(\sigma)$ is the spin part of the wave function. This prescription yielded a simple one-dimensional Schr\"{o}dinger equation for the remaining part of the wave function as,
\begin{equation}
\left[-\frac{\hbar^2}{2M}\frac{d^2}{dz^2} + V_{\textrm{eff,0m}}(z) - E_{m\nu}\right]Z_{m\nu}(z)=0,
\end{equation}
where the effective potential has the form,
\begin{eqnarray}
V_{\textrm{eff,0m}}(z) = -\frac{Ze^2}{a_H}\sqrt{\pi/2}\frac{(-1)^{|m|}}{|m|!}
\left(\frac{d}{d\kappa}\right)^{|m|} \nonumber\\
\times
\left[ 
\frac{1}{\sqrt{\kappa}}\exp \left (\frac{\kappa z^2}{2a_H^2}\right )\textrm{erfc} \left (\frac{\sqrt{\kappa}|z|}{\sqrt{2}a_H} \right)
\right]_{\kappa=1} \approx - \frac{Ze^2}{|z|+k_m a_H}.
\end{eqnarray}
Here 
\begin{equation}
k_m = \sqrt{\frac{2}{\pi}} \frac{2^{|m|} |m|!}{(2|m|-1)!!},
\end{equation}
with the double factorial begin defined by $(-1)!!=1$ and $(2n+1)!! = (2n+1)(2n-1)!!$. The approximate potential was designed to be valid to within 30\% over the entire domain with the explicit property that for large $m$, $\frac{1}{2}k_m a_H$ asymptotically approaches $\sqrt{2|m|+1}a_H$; the mean size of the Landau orbital. This simplification makes the problem analytically tractable resulting in Whittaker functions for the solution of the wave function along the $z$-direction. They additionally solved the problem numerically by alternatively expressing the $Z_{m\nu}(z)$ expanded using Gauss-Hermite functions (\emph{i.e.} the harmonic oscillator wavefunctions) as a basis set,
\begin{equation}
Z_{m\nu}(z)=\sum_{k=0}^{\infty} \frac{1}{(2\pi)^{1/4}\sqrt{a_Z 2^k k!}}A_{\nu m k} H_k\left (\frac{z}{\sqrt{2}a_Z}\right)\exp \left(-\frac{z^2}{4a_Z^2}\right),
\end{equation}
where $H_k(z)$ are the Hermite polynomials. It was seen that such a basis preserved the natural symmetries of the potential and consequently, with only a handful of basis functions it was possible to compute very accurately the binding energies of atoms and molecules in the intense field regime. A key enabling advantage of utilizing this basis within the Hartree-Fock method was that the computational overhead was significantly reduced in comparison to QMC and two-dimensional methods.

\begin{figure}[H]
\begin{center}
\includegraphics[width=\textwidth]{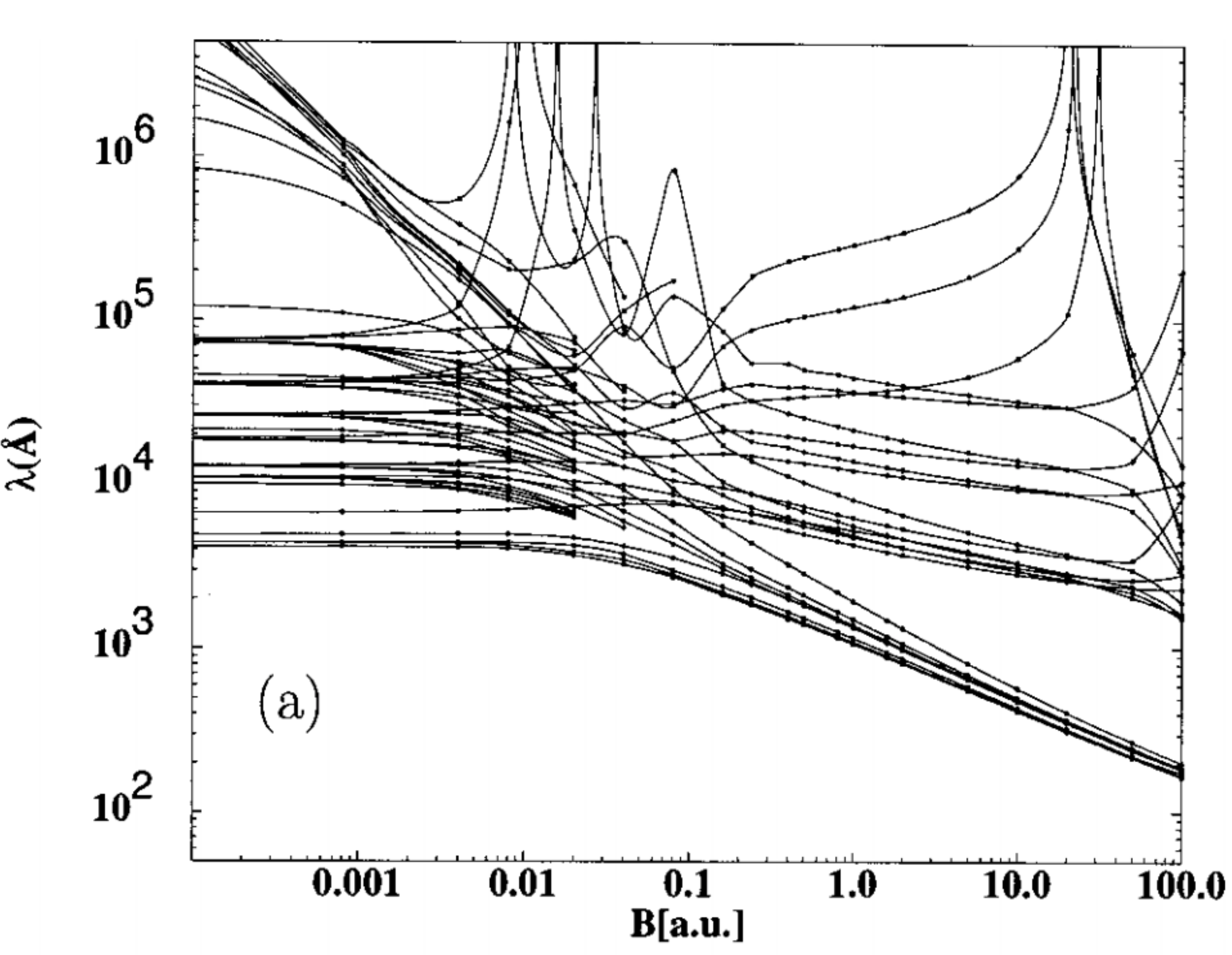}
\end{center}
\caption{Transition wavelengths in the singlet $\Delta M =1$ transitions as a function of magnetic field strength $B$ measured in atomic units Figure reprinted with permission from~\citet{Schmelcher2001}. Copyright 1994 by the American Physical Society.}
\label{fig:schmelcher2}
\end{figure}

Towards the end of the $1990$'s, Schmelcher and co-workers \citep{Schmelcher1988, Schmelcher1999, Schmelcher2000, Schmelcher2001, Schmelcher2003} developed a fully-correlated two-particle basis set which could be utilized over the entire range of magnetic field strengths ranging from weak to intense. The position representation of each individual electron's wave function was taken to have the explicit form,
\begin{equation}
\Phi_i(\rho,z,\phi) = \rho^{n_{\rho_i}} z^{n_{z_i}} e^{-\alpha_i \rho^2 - \beta_i z^2} e^{\textrm{i} m_i \phi},
\end{equation}
where $\alpha_i$ and $\beta_i$ are positive variational parameters and the exponents $n_{\rho_i}$ and $n_{z_i}$ obey the relationships,
\begin{equation}
n_{\rho_i} = |m_i|+2k_i~;~~~k_i=0,1,2,...~~~ \textrm{with}~~m_i=...,-2,-1,0,1,2,...
\end{equation}
\begin{equation}
n_{z_i} = \pi_{zi}+2l_i~;~~~l_i=0,1,2,...~~~ \textrm{with}~~\pi_{zi}=0,1
\end{equation}
The parameters $\alpha_i$ and $\beta_i$ were prescribed carefully chosen values which allows the wave function to be applicable to a whole range of magnetic field strengths. The Gaussian-like $\rho$-dependence of the wave function is similar to the ground Landau state, while the monomials $\rho^{n_{\rho_i}}$ and $z^{n_{z_i}}$ were tailored to be suitable for excitations. Their calculations were carried out using a configuration interaction formulation. This was a landmark development, as until then, many of the studies lost accuracy in different regimes depending upon the expansions employed, and in addition electron correlation which can account for an appreciable $1-2\%$ difference from Hartree-Fock estimates, had not been satisfactorily handled in the case of atomic structure in strong and intense magnetic fields. The accuracy of the work of Schmelcher et al is remarkable given that these calculations while still being computationally intensive due to the large number of configurations employed, were carried out with computing architectures prevalent in the late-$1990$'s and early $2000$'s, and to the current day their estimates for the binding energies of the various states of helium (and few-electron atoms) remain as a standard reference. Figure~\ref{fig:schmelcher2} shows the dependence of transition wavelengths for helium singlet transitions, as obtained by \citet[][]{Schmelcher2001}. 
Their studies also revealed that effects of electron correlation are still important in intense magnetic fields, despite the fact that the predominant interaction is with the magnetic field. They also addressed the important question of finite nuclear mass effects which become appreciable in intense magnetic fields. They were able to derive scaling formulae which enabled determination of the magnitude of this correction, based upon calculations for binding energies with infinite nuclear mass at certain scaled values of the magnetic field strength;
\begin{equation}
UH(M_0,B)U^{-1} = \mu \cdot H(\infty,B/\mu^2) - \frac{1}{M_0} B \cdot \sum_i (l_i + s_i),
\end{equation}
where, $M_0$ is the finite nuclear mass, $\mu = M_0/(1+M_0)$ is the reduced mass, and the unitary operator is given by $U=e^{-\textrm{i} \frac{1}{2} \textrm{ln} \mu(xp+px) }$.

\begin{figure}[H]
\begin{center}
\includegraphics[width=\textwidth]{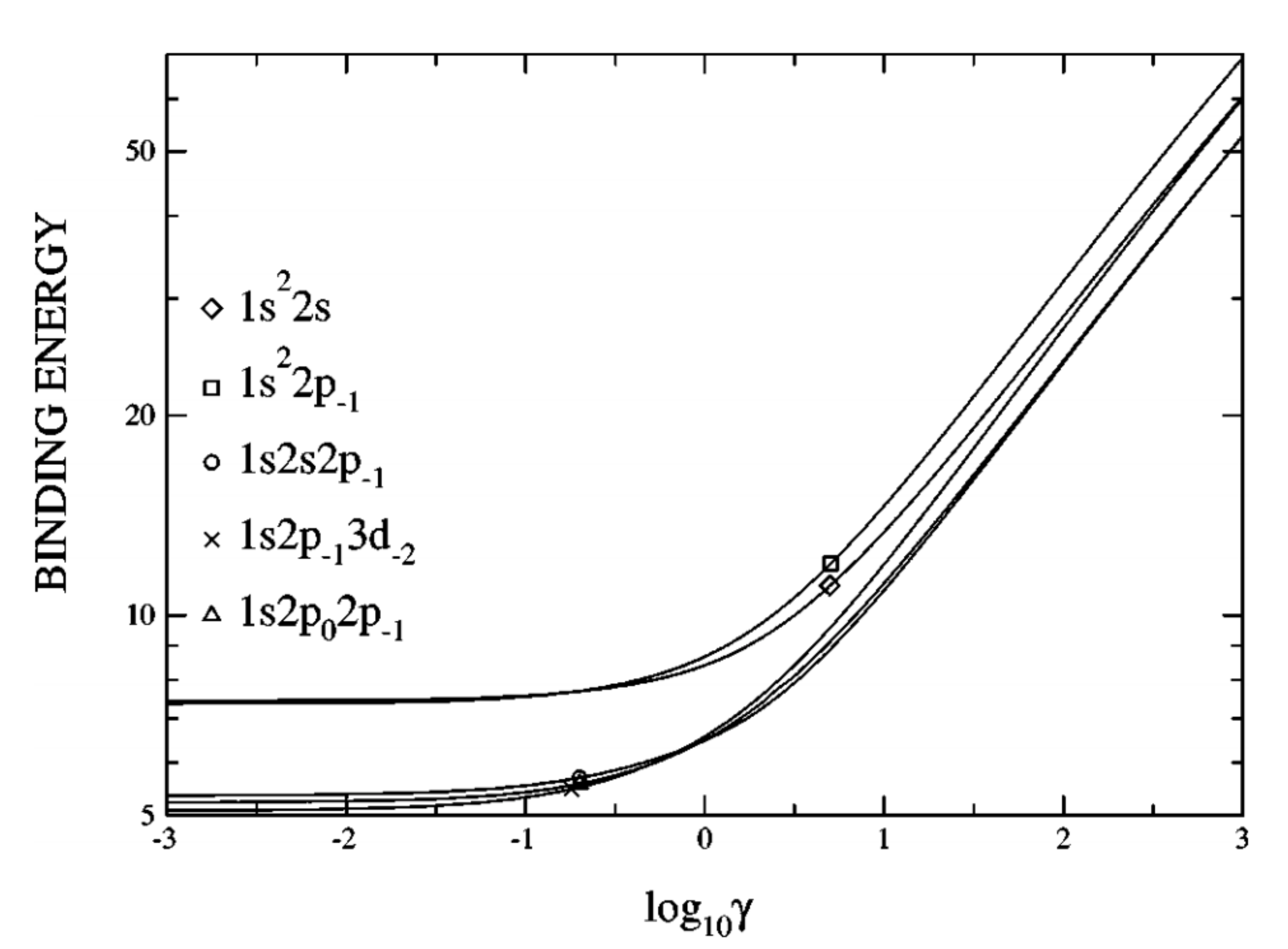}
\end{center}
\caption{Variation in the binding energy, measured in atomic units, of low-lying states of lithium with changing magnetic field strength.  Figure reprinted with permission from~\citet{Ivanov1997}. Copyright 1998 by the  American Physical Society. }
\label{fig:ivanov_lithium_energy}
\end{figure}

\begin{figure}[H]
\begin{center}
\includegraphics[width=\textwidth]{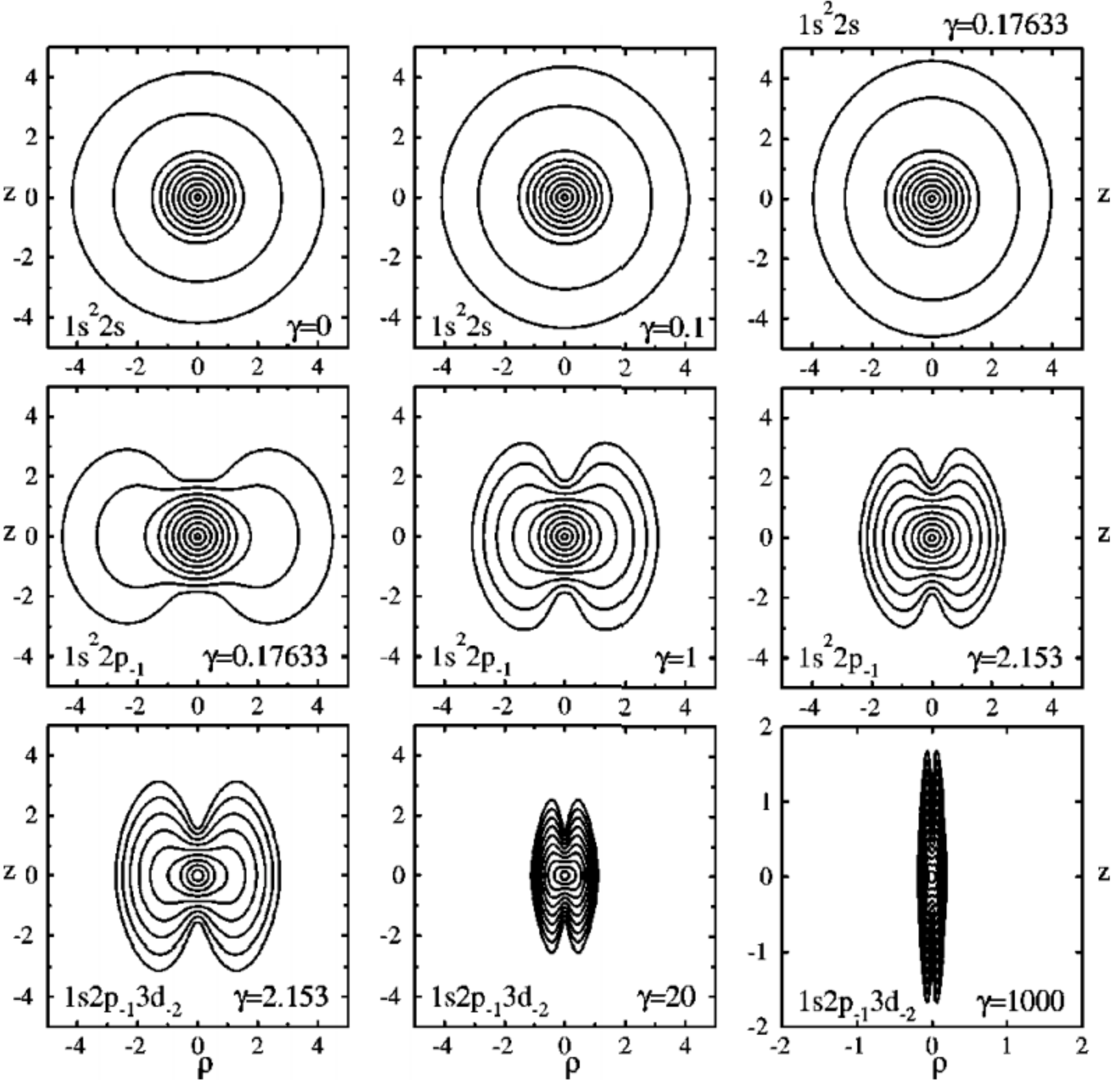}
\end{center}
\caption{Electron densities of a few low-lying states of lithium in strong to intense magnetic fields, measured by the $\gamma$ parameter.  Figure reprinted with permission from~\citet{Ivanov1997}. Copyright 1998 by the  American Physical Society.}
\label{fig:ivanov_lithium_wavefuncs}
\end{figure}

\begin{figure}
\begin{center}
\includegraphics[width=12 cm]{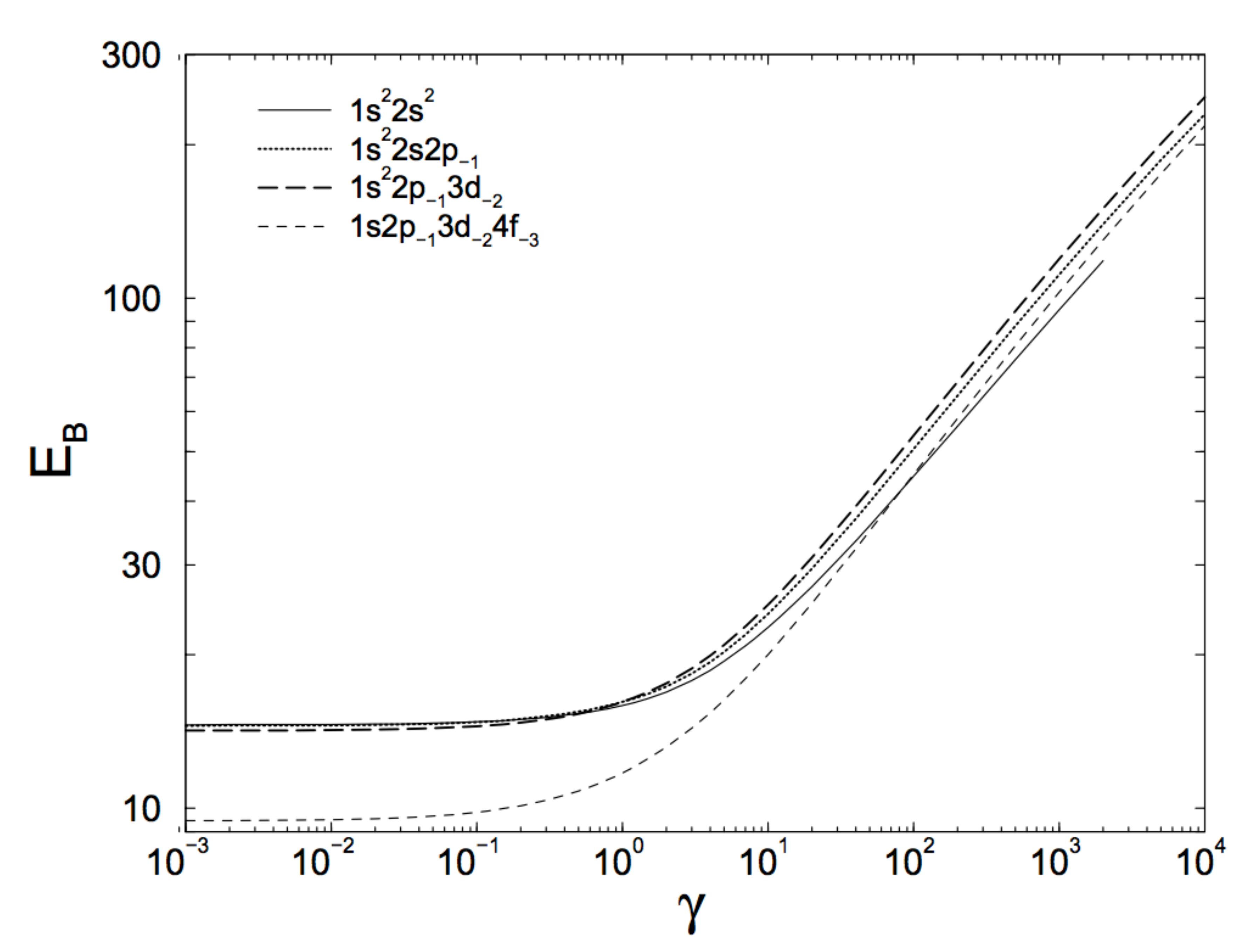}
\end{center}
\caption{The binding energies (in atomic units) of the ground
state electronic configurations of the Be atom depending on
the magnetic field strength. The field strength is given in units
of $ \gamma = (B/B_0), B_0 = \hbar c/ea_0^2 = 2.3505 \times 10^5$~T. Reprinted from European Journal of Physics, D, volume 14, 2001, 270-288 ``The beryllium atom and beryllium positive ion in strong magnetic fields", M.V.~Ivanov and P.~Schmelcher, Figure 4, copyright 2001 Springer. Figure reprinted with kind permission from Springer Science and Business Media.}
\label{fig:ivanov_beryllium_energy}
\end{figure}

\begin{figure}[h]
\begin{center}
\includegraphics[width=12 cm]{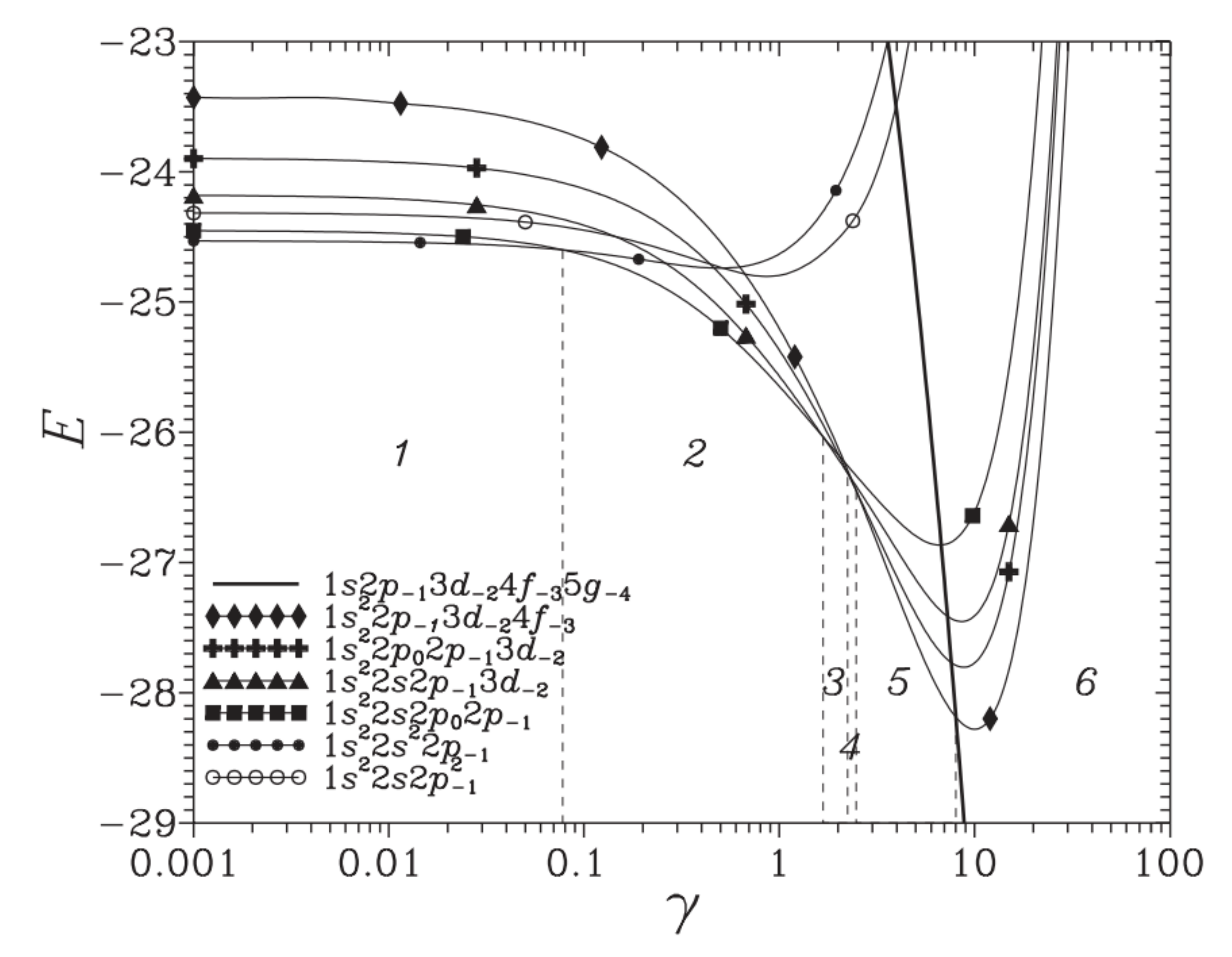}
\end{center}
\caption{Binding energy, in atomic units of low-lying states of boron as a function of magnetic field strength. Figure reprinted with permission from~\citet{Schmelcher_boron2001}. Copyright 2001 by the IOP Publishing. Reproduced with permission. All rights reserved.}
\label{fig:ivanov_boron_energy}
\end{figure}

Elsewhere,  \citet{IS2000_2} carried out two-dimensional Hartree-Fock calculations for determining the ground state energies of atoms up to $Z=10$, in strong and intense magnetic fields. These however did not include effects of electron correlation. They determined the ground state energies of these atoms by looking at both the fully and partially spin-polarized states. In the former all the electron spins are anti-aligned with the magnetic field to minimize energy, while in the latter only some of the electrons are anti-aligned. Typically the former type of states are favored in intense magnetic fields as they become more tightly bound. In their study they were also able to determine ground state cross-overs. Over the next few years, Schmelcher and Ivanov et al systematically investigated using both simple Hartree-Fock as well as configuration interaction calculations, the first few low-lying states of atoms such as lithium \citep{Ivanov1997, Schmelcher_lithium2004}, beryllium \citep{Ivanov2001EPJD, Schmelcher_beryllium2004}, boron \citep{Schmelcher_boron2001} and carbon \citep{Ivanov1999}, in strong and intense magnetic fields and these studies represent nearly all of the data that is available in the literature for the structure of light atoms in strong (and intense) magnetic fields. Figure~\ref{fig:ivanov_lithium_energy} shows the variation in the binding energy of low-lying states of lithium with magnetic field strength while Figure~\ref{fig:ivanov_lithium_wavefuncs} shows how the wave functions of the low-lying states of lithium change with increasing magnetic field strength. 
Similarly, Figure~\ref{fig:ivanov_beryllium_energy} and \ref{fig:ivanov_boron_energy} show the binding energies of beryllium and boron as functions of magnetic field strength, respectively.

Elsewhere, Medin and Lai investigated atoms and molecules \cite[][]{Medin2006I} as well as chains of atoms and molecules \cite[][]{Medin2006II} in strong and intense magnetic fields using DFT. However they were only able to investigate the ground state of atoms such as helium, carbon and iron. Their motivation was more with regards to investigating properties of the solid crusts of neutron stars. A year later, B\"ucheler et al \cite[][]{Bucheler2007, Bucheler2008} were able to apply the method of released-phase QMC to study the ground states of atoms up to $Z=26$ at a magnetic field strength of $5 \times 10^{12}$~G. This represents one of literally a handful of investigations for accurate data for the ground states of many of these atoms in an intense magnetic field. Even then quite crucially, data is not available for other states or for other magnetic field strengths. Elsewhere, Engel and Schimeczek and co-workers, working with G\"unter Wunner, investigated atoms in strong and intense magnetic fields using two separate approaches. First, they carried out fixed-phase QMC calculations and arrived at estimates for the ground states of atoms from $Z=2$ to $26$ \cite[][]{Meyer2013} as well as a Hartree-Fock-Roothan method with a fast parallel implementation using finite-element techniques \cite[][]{Schimeczek2012, Engel2008, Engel2009}, in all cases obtaining beautifully accurate results for the ground states of atoms as well as for oscillator strengths. They expand the wave function as
\begin{equation}
\psi^i(\rho_i,z_i,\phi_i)=\sum_{n=0}^{N_L} \sum_{\nu} \alpha^i_{n\nu}B^i_{\nu}(z_i) \Phi_{nm_i}(\rho_i,\phi_i),
\end{equation}
where the $z$-dependence of the expansion has been expanded in terms of a B-spline basis of functions. They consider up to $N_L$ different Landau channels with a different unknown $z$-part of the wave function in each channel. Utilizing Landau levels for two of the three orthogonal directions ($\{\rho,\phi \}$), simplifies their eigenvalue computation significantly and this allows them to solve the one-dimensional problem of determining the unknown $z$-component of the wave functions, in a highly economical way. Recently, the authors of the current article also investigated the lithium atom in strong and intense magnetic fields using a fully two-dimensional pseudospectral Hartree-Fock method \cite[][]{HT2010, Thirumalai2012}, obtaining data for both the ground and some other low-lying states of the lithium atom that have not been investigated thus far in the literature. The hallmark of these methods is that the computation time is greatly reduced, despite the fact that the problem is fully two-dimensional, chiefly by virtue of spectral convergence. Computational times are reduced to a matter of mere seconds for obtaining accurate data for the binding energies making such implementations highly desirable for ease of integration with atmosphere models of neutron stars and white dwarfs. Table~\ref{tab:table1} shows data for two hitherto un-calculated states of the lithium atom from such a calculation \cite[][]{Thirumalai2012}.  
\begin{table}[H]
\centering
\caption{Binding energies of two hitherto un-calculated negative parity states of lithium, due to \citet[][]{Thirumalai2012}.}
\begin{tabular}{c@{\hspace{3mm}}c@{\hspace{3mm}}c@{\hspace{3mm}}}
\hline
\hline
 $\beta_Z$ & $1^4(-2)^-$ & $1^4(-3)^-$  \\
\hline
 1 	 	&  3.0074   	& 2.9807 	\\
10 	 	& 6.6313   	& 6.6095 	\\
50 	 	& 11.3941 	& 11.3809	\\
100 	 	& 14.2445 	& 14.2331	\\
200 	 	& 17.6795 	& 17.6601	\\
500 	 	& 23.2339 	& 23.2195	\\
1000 	& 28.3062 	& 28.2948 \\
\hline 
\hline
\end{tabular}
\label{tab:table1}
\end{table}

\section{\label{sec:Conc}Concluding remarks and future prospects}
Presently we could be said to be in the post-modern era for atomic structure calculations, with large scale computational capabilities at our disposal. The state-of-the-art computing facilities today boast of petaflop processors with terabytes of computer memory available for computations. The problem of atomic structure in strong magnetic fields today is primarily a computational one, with efforts in two simultaneous directions. First, trying to determine the spectrum of low-lying states of low-$Z$ atoms that have not been investigated so far, and second improving the estimates of the currently determined binding energies and oscillator strengths using post-HF techniques. Both these avenues require computing resources which are becoming available today. As spectrometers become more sensitive, data will begin to emerge for the spectra of neutron stars. At which point, for interpreting the spectra, researchers will not only need data for many of the states of atoms in intense magnetic fields, they will also need highly accurate data for oscillator strengths and bound-bound and bound-free transitions. They will also need extensive data for atoms in crossed electric and magnetic fields, which will drastically alter the spectrum; such strong electric fields can exist in the plasma in the atmospheres of neutron stars. Aside from the motivation to analyze spectra, the fundamental question, ``what do different atoms in the periodic table look like in strong magnetic fields?" is, as of the writing of this article, a largely uncharted domain, where we only understand well the two most basic atoms of the universe; hydrogen and helium. After well over a century since Zeeman's original discovery, we are still trying to answer this fundamental question with regard to low-$Z$ atoms. The current era is an exciting one for light atoms in strong magnetic fields, primarily due to advances in computing and numerical techniques, and it is the hope of the authors that soon these problems, which are currently active fields of research, will be relegated to the pages of textbooks, under the category of ``solved problems".

\bibliographystyle{advamop}
\bibliography{articleEA}

\end{document}